\documentstyle[preprint,aps,epsfig,epsf,floats,tighten]{revtex}
\newcommand{\MET}{\hbox{$\rlap{\kern0.25em/}E_T$}}
\newcommand{\MEX}{\hbox{$\rlap{\kern0.25em/}E_x$}}
\newcommand{\MEY}{\hbox{$\rlap{\kern0.25em/}E_y$}}
\newcommand{\METcal}{\hbox{$\rlap{\kern0.25em/}E_T^{{\rm cal}}$}}
\begin{document}
\title{Limits on Anomalous $WW\gamma$ and $WWZ$ Couplings from
$WW/WZ\rightarrow e\nu jj$ Production}

%
\author{                                                                      
B.~Abbott,$^{47}$                                                             
M.~Abolins,$^{44}$                                                            
V.~Abramov,$^{19}$                                                            
B.S.~Acharya,$^{13}$                                                          
D.L.~Adams,$^{54}$                                                            
M.~Adams,$^{30}$                                                              
S.~Ahn,$^{29}$                                                                
V.~Akimov,$^{17}$                                                             
G.A.~Alves,$^{2}$                                                             
N.~Amos,$^{43}$
% ...
\\                                                                            
\vskip 0.30cm                                                                 
\centerline{(D\O\ Collaboration)}                                             
\vskip 0.30cm                                                                 
}                                                                             
\address{                                                                     
\centerline{$^{1}$Universidad de Buenos Aires, Buenos Aires, Argentina}       
\centerline{$^{2}$LAFEX, Centro Brasileiro de Pesquisas F{\'\i}sicas,         
                  Rio de Janeiro, Brazil}                                     
\centerline{$^{3}$Universidade do Estado do Rio de Janeiro,                   
                  Rio de Janeiro, Brazil}                                     
\centerline{$^{4}$Institute of High Energy Physics, Beijing,                  
                  People's Republic of China}                                 
\centerline{$^{5}$Universidad de los Andes, Bogot\'{a}, Colombia}             
\centerline{$^{6}$Universidad San Francisco de Quito, Quito, Ecuador}         
\centerline{$^{7}$Institut des Sciences Nucl\'eaires, IN2P3-CNRS,             
                  Universite de Grenoble 1, Grenoble, France}                 
\centerline{$^{8}$Centre de Physique des Particules de Marseille,             
                  IN2P3-CNRS, Marseille, France}                              
\centerline{$^{9}$LPNHE, Universit\'es Paris VI and VII, IN2P3-CNRS,          
                  Paris, France}                                              
\centerline{$^{10}$DAPNIA/Service de Physique des Particules, CEA, Saclay,    
                  France}                                                     
\centerline{$^{11}$Panjab University, Chandigarh, India}                      
\centerline{$^{12}$Delhi University, Delhi, India}                            
\centerline{$^{13}$Tata Institute of Fundamental Research, Mumbai, India}     
\centerline{$^{14}$Seoul National University, Seoul, Korea}                   
\centerline{$^{15}$CINVESTAV, Mexico City, Mexico}                            
\centerline{$^{16}$Institute of Nuclear Physics, Krak\'ow, Poland}            
\centerline{$^{17}$Institute for Theoretical and Experimental Physics,        
                   Moscow, Russia}                                            
\centerline{$^{18}$Moscow State University, Moscow, Russia}                   
\centerline{$^{19}$Institute for High Energy Physics, Protvino, Russia}       
\centerline{$^{20}$Lancaster University, Lancaster, United Kingdom}           
\centerline{$^{21}$University of Arizona, Tucson, Arizona 85721}              
\centerline{$^{22}$Lawrence Berkeley National Laboratory and University of    
                   California, Berkeley, California 94720}                    
\centerline{$^{23}$University of California, Davis, California 95616}         
\centerline{$^{24}$California State University, Fresno, California 93740}     
\centerline{$^{25}$University of California, Irvine, California 92697}        
\centerline{$^{26}$University of California, Riverside, California 92521}     
\centerline{$^{27}$Florida State University, Tallahassee, Florida 32306}      
\centerline{$^{28}$University of Hawaii, Honolulu, Hawaii 96822}              
\centerline{$^{29}$Fermi National Accelerator Laboratory, Batavia,            
                   Illinois 60510}                                            
\centerline{$^{30}$University of Illinois at Chicago, Chicago,                
                   Illinois 60607}                                            
\centerline{$^{31}$Northern Illinois University, DeKalb, Illinois 60115}      
\centerline{$^{32}$Northwestern University, Evanston, Illinois 60208}         
\centerline{$^{33}$Indiana University, Bloomington, Indiana 47405}            
\centerline{$^{34}$University of Notre Dame, Notre Dame, Indiana 46556}       
\centerline{$^{35}$Purdue University, West Lafayette, Indiana 47907}          
\centerline{$^{36}$Iowa State University, Ames, Iowa 50011}                   
\centerline{$^{37}$University of Kansas, Lawrence, Kansas 66045}              
\centerline{$^{38}$Kansas State University, Manhattan, Kansas 66506}          
\centerline{$^{39}$Louisiana Tech University, Ruston, Louisiana 71272}        
\centerline{$^{40}$University of Maryland, College Park, Maryland 20742}      
\centerline{$^{41}$Boston University, Boston, Massachusetts 02215}            
\centerline{$^{42}$Northeastern University, Boston, Massachusetts 02115}      
\centerline{$^{43}$University of Michigan, Ann Arbor, Michigan 48109}         
\centerline{$^{44}$Michigan State University, East Lansing, Michigan 48824}   
\centerline{$^{45}$University of Nebraska, Lincoln, Nebraska 68588}           
\centerline{$^{46}$Columbia University, New York, New York 10027}             
\centerline{$^{47}$New York University, New York, New York 10003}             
\centerline{$^{48}$University of Rochester, Rochester, New York 14627}        
\centerline{$^{49}$State University of New York, Stony Brook,                 
                   New York 11794}                                            
\centerline{$^{50}$Brookhaven National Laboratory, Upton, New York 11973}     
\centerline{$^{51}$Langston University, Langston, Oklahoma 73050}             
\centerline{$^{52}$University of Oklahoma, Norman, Oklahoma 73019}            
\centerline{$^{53}$Brown University, Providence, Rhode Island 02912}          
\centerline{$^{54}$University of Texas, Arlington, Texas 76019}               
\centerline{$^{55}$Texas A\&M University, College Station, Texas 77843}       
\centerline{$^{56}$Rice University, Houston, Texas 77005}                     
\centerline{$^{57}$University of Washington, Seattle, Washington 98195}       
}                                                                             

\maketitle

\date{\today}

\begin{abstract}
Limits on anomalous $WW\gamma$ and $WWZ$ couplings are presented from a study of
$WW/WZ\rightarrow e\nu jj$ events produced in $p{\bar p}$
collisions at $\sqrt{s}=1.8$ TeV. 
Results from the analysis of data collected using the D{\O}
detector during the 1993--1995 Tevatron collider run at Fermilab
are combined with those of an earlier study from
the 1992--1993 run.
A fit to the transverse momentum spectrum of the $W$ boson yields direct limits
on anomalous $WW\gamma$ and $WWZ$ couplings.
With the assumption that the $WW\gamma$ and $WWZ$ couplings are equal, we
obtain $-0.34< \lambda < 0.36$ (with $\Delta\kappa=0$) and
$-0.43< \Delta\kappa < 0.59$ (with $\lambda=0$) at the 95\% confidence level
for a form-factor scale $\Lambda=2.0$ TeV.
\end{abstract}

\pacs{PACS numbers: 14.70.-e 12.15.Ji 13.40.Em 13.40.Gp }


\section{Introduction}
\label{introduction}

\begin{figure}[htb]
\epsfxsize=3.in
\centerline{\epsffile{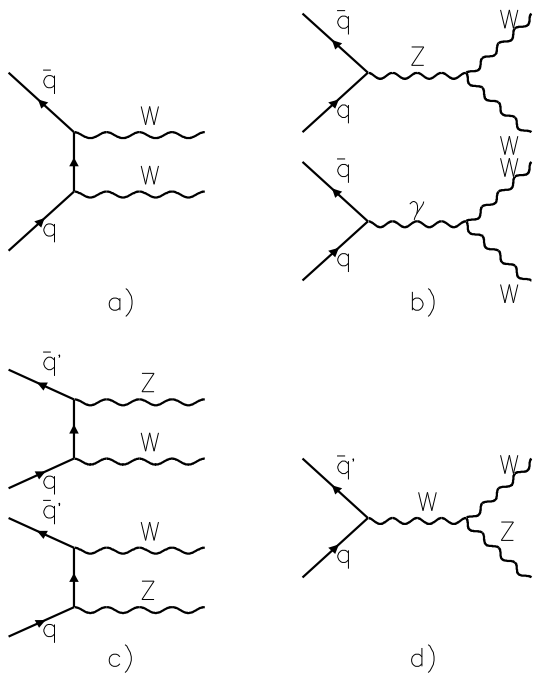}}  
\caption{Feynman diagrams for $WW$ and $WZ$ production at leading order. a) and
c): $t$- and $u$-channel quark exchange diagrams; b) and d): $s$-channel
diagrams with three-boson couplings.}
\label{fig:feynman}
\end{figure}

	The Tevatron $p{\bar p}$ collider at Fermilab offers one of
the best opportunities to test trilinear gauge boson
couplings~\cite{Baur90,Hagiwara90,Ellison}, which are a direct
consequence of the non-Abelian $SU(2)\times U(1)$
gauge structure of the standard model (SM). The trilinear gauge boson
couplings can be measured directly from gauge boson pair (diboson) 
production.
Production of $WW$ and $WZ$ pairs in
$p\bar{p}$ collisions at $\sqrt{s} = 1.8$ TeV can proceed through $s$-channel
boson intermediaries, or a $t$- or $u$-channel quark exchange processes as
shown in Fig.~\ref{fig:feynman}.
There are important cancellations between
the $t$- or $u$-diagrams, which involve only couplings of the bosons
to fermions, and the $s$-channel diagrams which contain three-boson
couplings.  
These cancellations are essential for making calculations of SM diboson
production unitary and renormalizable.
Since the fermionic couplings of the $\gamma$ and $W$ and $Z$ bosons 
have been well tested~\cite{PDG}, we may regard diboson production as
primarily a test of the three-boson vertex.
Production of $WW$ pairs is sensitive to both $WW\gamma$ and $WWZ$ couplings; 
$WZ$ production is sensitive only to $WWZ$ couplings. 

A generalized effective
Lagrangian has been developed to describe the couplings of three
gauge bosons~\cite{Hagiwara87}. The
Lorentz-invariant effective Lagrangian for the gauge boson self-interactions
contains fourteen dimensionless coupling parameters,
$\lambda_V$, $\kappa_V$, $g_1^V$, $\tilde{\lambda}_V$, $\tilde{\kappa}_V$,
$g_4^V$, and $g_5^V$ ($V=Z$ or $\gamma$), seven for
$WWZ$ interactions and another seven for $WW\gamma$ interactions, and two
overall couplings, $g_{WW\gamma} = - e$ and $g_{WWZ}= -e \cot \theta_{W}$,
where $e$ and $\theta_W$ are the positron charge and the weak
mixing angle. The couplings $\lambda_V$ and $\kappa_V$ conserve charge $C$ and
parity $P$. The couplings $g_4^V$ are odd under $CP$ and $C$, $g_5^V$ are odd
under $C$ and $P$, and $\tilde{\kappa}_V$ and $\tilde{\lambda}_V$ are odd
under $CP$ and $P$. 
To first order in the SM (tree level), all of the couplings vanish
except $g_1^V$ and $\kappa_V$ ($g_1^{\gamma} = g_1^Z =
\kappa_{\gamma} = \kappa_Z = 1$).
For real photons,
gauge invariance in electromagnetic interactions does not
allow deviations of $g_1^{\gamma}$, $g_4^{\gamma}$, and $g_5^{\gamma}$ from
their SM values of 1, 0, and 0, respectively. The
$CP$-violating $WW\gamma$ couplings $\tilde{\lambda}_{\gamma}$ and
$\tilde{\kappa}_{\gamma}$ are tightly constrained by measurements of the
neutron electric dipole moment~\cite{Boudjema}. In the present study, we assume
$C$, $P$ and $CP$ symmetries are conserved, reducing the independent coupling
parameters to $\kappa_{\gamma}$, $\kappa_Z$, $\lambda_{\gamma}$,
$\lambda_Z$ and $g_1^Z$. 

Cross sections for gauge boson pair production increase
for couplings with non-SM values,
because the cancellation between the $t$- and $u$-channel diagrams and
the $s$-channel diagrams is destroyed.
This can yield
large cross sections at high energies, eventually
violating tree-level
unitarity. A consistent description therefore requires anomalous couplings
with a form factor that causes them to vanish at
very high energies.  We will use dipole form factors, e.g., 
$\lambda_V(\hat{s}) = \lambda_V/(1 + \hat{s}/\Lambda^2)^2$, where $\hat{s}$
is the square of the invariant mass of the gauge-boson pair.
Given a form-factor scale $\Lambda$, the anomalous-coupling
parameters are restricted by $S$-matrix unitarity.
Assuming that the independent coupling parameters
are $\kappa=\kappa_{\gamma}=\kappa_Z$ and
$\lambda=\lambda_{\gamma}=\lambda_Z$, tree-level unitarity is
satisfied if
$\Lambda \leq [ 6.88/((\kappa - 1)^2 + 2\lambda^2] ^{1/4}$
TeV~\cite{Hagiwara90,Baur}.
The experimental limits on anomalous couplings can be compared with
the bounds derived from $S$-matrix unitarity, and constrain the
trilinear gauge-boson couplings only if the limits are more
stringent than the bounds from unitarity for any given value of $\Lambda$.
 
For both $WW$ and $WZ$ production processes,
the effect of anomalous values of $\lambda_V$ on the helicity amplitudes
is enhanced for large $\hat{s}$.
On the other hand, terms containing $\Delta\kappa_{V}$ $(=\kappa_{V} - 1)$
grow as $\sqrt{\hat{s}}$ in the $WZ$ production process, and
as $\hat{s}$ in the $WW$ production process.
Limits on $\Delta\kappa_V$ from the study of $WW$
production are therefore expected to be tighter than those
from $WZ$ production. 

\begin{figure}[htb]
\epsfxsize=3.in
\centerline{\epsffile{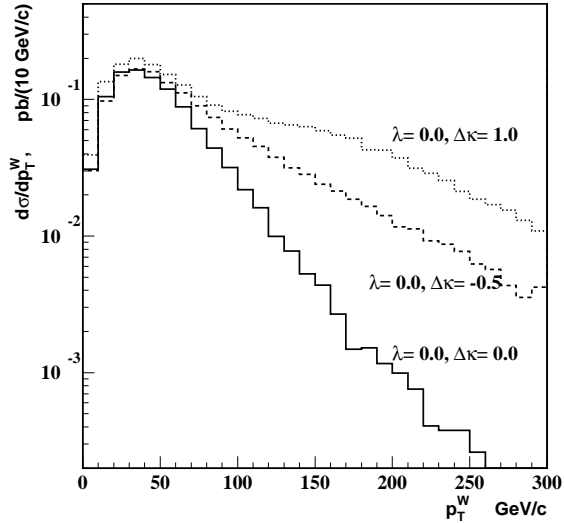}}
\caption{The $p_T^W$ spectrum of generated
$p\bar{p}\rightarrow WW\rightarrow e\nu jj$ events
with SM couplings and two examples of anomalous couplings.}
\label{fig:pt_anomalous}
\end{figure}

Since anomalous couplings contribute only via $s$-channel photon or $W$ or $Z$
boson
intermediaries, their effects are expected mainly in the region of small
vector boson rapidities, and the transverse momentum distribution of the
vector boson is therefore particularly sensitive to anomalous trilinear
gauge-boson couplings. 
This is demonstrated in Fig.~\ref{fig:pt_anomalous}, which shows the
distribution of the $W$ boson transverse momentum $p_T^W$ in simulated
$p\bar{p} \rightarrow WW+X\rightarrow e\nu jj+X$ events
for anomalous trilinear gauge boson couplings, using a
dipole form factor with a scale $\Lambda = 1.5$ TeV, and with the couplings
for $WW\gamma$ and $WWZ$ assumed to be equal.

	Trilinear gauge-boson couplings can therefore be measured by
comparing the shapes of the $p_T$ distributions of
the final state gauge bosons with theoretical predictions. 
Even if the background is much larger than the expected gauge-boson pair
production
signal as is the case for the $WW/WZ\rightarrow e\nu jj$ process, limits on
anomalous couplings can still be set using a kinematic region
where the effects of anomalous trilinear gauge boson couplings
are expected to dominate.

	Trilinear gauge-boson couplings have been studied in several 
experiments. 
$WW\gamma$ couplings have been studied
in $p\bar{p}$ collisions by the UA2~\cite{UA2}, CDF~\cite{CDFWg}, and
D{\O}~\cite{D0Wg,D01aPRD} collaborations using $W\gamma$ events.
The UA2 results are based on
data taken during the 1988--1990 CERN $p\bar{p}$ collider run at $\sqrt{s}
= 630$ GeV with an integrated luminosity of 13 pb$^{-1}$ and
the CDF and D{\O} data are from the 1992--1993 and 1993--1995 Fermilab
$p\bar{p}$ runs at $\sqrt{s} = 1.8$ TeV. $WWZ$ couplings together
with the $WW\gamma$ couplings have been studied by the CDF
and D{\O} collaborations using $W$ boson pair production in the dilepton decay
modes~\cite{D01aPRD,CDFWW,D0WW} and $WW$/$WZ$ production in the single-lepton
modes~\cite{D01aPRD,CDFWZ,D0WZ1a,D0WZ1bmu}.
Experiments at the CERN LEP Collider have recently
reported results of similar studies~\cite{LEP}.

In this report, we present a detailed description of previously summarized 
work~\cite{D0WZ1b} on
$WW$ and $WZ$ production with one $W$ boson decaying into an electron
(or a positron) and an antineutrino (or a neutrino) and a second $W$
or $Z$ boson decaying into two jets~\cite{Sanchez-thesis}.
Due to the limitation in  jet-energy resolution, 
the hadronic decay of a $W$ boson can not be differentiated from that of
a $Z$ boson.
This analysis is based on the data collected during the 1993--1995 Tevatron
collider run at Fermilab.
From the observed candidate events and
background estimates, 95\% confidence level (C.L.) limits are set on the
anomalous trilinear gauge boson couplings.
The results are combined with those from the 1992--1993 data to provide the
final limits on the couplings from the D{\O} analysis.

Brief summaries of the detector and the
multilevel trigger and data acquisition systems are presented in
Sections~\ref{detector} and \ref{trigger-daq}.
Sections~\ref{particle-id}, \ref{data-sample} and \ref{event-selection}
describe our particle identification methods, the data sample,
and event selection criteria.
Sections~\ref{detection-efficiency} and \ref{background} are
devoted to detection efficiency and background
estimates.
Results and conclusions are presented in Sections~\ref{results} and
\ref{conclusions}.

\section{The D{\O} Detector}
\label{detector}

\begin{figure}[htb]
\epsfxsize=3.4in
\centerline{\epsffile{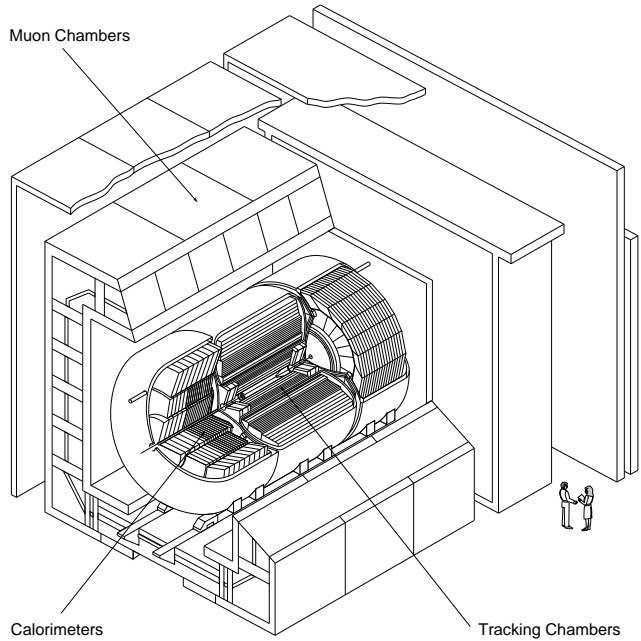}} 
\caption{Cutaway view of the D{\O} detector}
\label{fig:d0}
\end{figure}

    The D{\O} detector~\cite{D0detector}, illustrated in Fig.~\ref{fig:d0},
is a general-purpose detector designed for the
study of proton-antiproton collisions at $\sqrt{s}$ = 1.8 TeV and
is located at the D{\O} interaction region of the Tevatron ring at Fermilab. 

The innermost part of the detector consists of a set
of tracking chambers that surround the beam pipe.
There is no central magnetic field and jets are measured using a compact
set of calorimeters positioned outside the tracking volume.
To identify muons, an additional set
of tracking chambers is located outside the calorimeter, with
a measurement of muon momentum provided through magnetized iron toroids
placed between the first two muon-tracking layers. 

The full detector is about 13 m high $\times$ 11 m wide $\times$ 17 m long,
with a total weight of about 5500 tons. The Tevatron beam pipe passes
through the center of the detector, while the Main Ring beam pipe
passes through the
upper portion of the calorimetry, approximately 2 m above the Tevatron beam
pipe. The coordinate system used in D{\O} is right-handed,
with the $z$-axis pointing along the 
direction of the proton beam (southward) and the $y$-axis pointing up.
The polar angle $\theta=0$ is along the proton beam direction,
and the azimuthal
angle $\phi=0$ along the eastward direction. Instead of $\theta$, 
we often use the
pseudorapidity, $\eta=-\ln[\tan(\theta/2)]$. This
quantity approximates the true rapidity $y = 1/2 \ln[(E+p_z)/(E-p_z)]$,
when the rest mass is much smaller than the total energy. 

\subsection{Central Detector}
\label{central-detector}

The tracking chambers and a transition radiation detector make up the central
detector (CD). The main purpose of the CD is to measure
the trajectories of charged particles and determine the $z$ position of
the interaction vertex. This information can be used to determine whether an
electromagnetic energy cluster in the calorimeter is
caused by an electron or by
a photon.  Additional information such as the number of tracks and
the ionization energy along the track ($dE/dx$) can be used to determine
whether
a track is caused by one or several closely spaced charged particles,
such as a photon conversion.

The CD consists of four separate subsystems: the vertex drift
chamber (VTX), the transition radiation detector (TRD), the central drift
chamber (CDC), and two forward drift chambers (FDC). The full set of CD
detectors fits within the inner cylindrical aperture of the calorimeters
in a volume of radius $r$ = 78 cm and length $l$ = 270 cm. The system
provides charged-particle tracking over the region $|\eta| <$~3.2.
The trajectories of charged particles are measured with a
resolution of 2.5 mrad in $\phi$ and 28 mrad in $\theta$. From these
measurements, the position of the interaction vertex 
along the $z$ direction is determined with a resolution of 6 mm. 

	The VTX is the innermost tracking chamber in the D{\O} detector,
occupying the region $r=3.7$ cm to 16.2 cm. It
is made of three mechanically independent concentric layers of cells
parallel to the beam pipe.   The innermost layer
has sixteen cells while the outer two layers have thirty-two cells each. 

	The TRD occupies the space between the VTX and the CDC; it extends from
$r=17.5$ cm to 49 cm.
The TRD consists of three separate units, each containing a radiator (393
foils of 18 $\mu$m thick polypropylene in a volume filled with nitrogen gas)
and an X-ray detection chamber filled with Xe gas.
The TRD information is not used in this analysis. 

	The CDC is a cylindrical drift chamber, 184 cm along $z$, located
between $r=49.5$ and $r=74.5$ cm,
and provides coverage for $|\eta| <$ 1.2. It is made
up of four concentric rings of 32 azimuthal cells per ring. Each cell
contains seven sense wires (staggered by 200 $\mu$m relative to each
other to help resolve left-right ambiguities), and two delay lines.
The $r\phi$ position of a hit is determined via the
drift time measured for the hit wire and the $z$ position of a hit is
measured using
inductive delay lines embedded in the module-walls of the sense wire
planes.

	The FDC consists of two sets of drift
chambers located at the ends of the CDC. They perform the same function as
the CDC, but for 1.4~$< |\eta| <$~3.1. Each FDC package consists of three
separate chambers: a $\Phi$ module, whose sense wires are radial and
measure the $\phi$ coordinate, sandwiched between a pair of $\Theta$
modules whose sense wires measure the $\theta$ coordinate. 

\subsection{Calorimeters}
\label{calorimeters} 

	The D{\O} calorimeters are sampling calorimeters, with liquid argon
as the sensitive ionization medium. The primary absorber material is depleted
uranium, with copper and stainless steel used in the outer regions. There
are three separate units, each contained in separate cryostats: 
the Central Calorimeter (CC), the North End Calorimeter (ECN), and the
South End Calorimeter (ECS). 
The readout cells are arranged in a pseudo-projective geometry pointing to
the interaction region.

The calorimeters are 
subdivided in depth into three distinct types of modules:
electromagnetic sections (EM) with relatively thin uranium absorber
plates, fine-hadronic sections (FH) with thicker uranium plates, and
coarse-hadronic sections (CH) with thick copper or stainless steel plates.
There are four separate layers for the EM modules in both the CC and EC
that are readout separately.
The first two layers are 2 radiation lengths thick in the CC and 0.3 and 2.6
radiation lengths thick in the EC, 
and measure the initial longitudinal shower development,
where photons and $\pi^0$s differ somewhat on a statistical basis. The
third layer spans the region of maximum EM shower energy deposition and the
fourth completes the EM coverage of approximately 20 total radiation lengths.
The fine-hadronic modules are typically segmented into three or four layers. 
Typical transverse sizes of towers in both EM and hadronic modules are
$\Delta \eta = 0.1$ and $\Delta \phi =2 \pi / 64 \approx 0.1$. The third
section of the EM modules is segmented twice as finely in both $\eta$ and
$\phi$ to provide more precise determination of centroids of EM showers. 

The CC has a length of 2.6 m, covering the pseudorapidity region $|\eta| <
$ 1.2, and consists of three concentric cylindrical rings.  There are 32 EM
modules in the inner ring, 16 FH modules in the
surrounding ring, and 16 CH modules in the outer ring. 
The EM, FH and CH module boundaries are rotated 
with respect to each other so as to prevent having
more than one intermodular gap intercepting a trajectory from the origin of
the detector.

The two end calorimeters (ECN and ECS) are mirror-images, and
contain four types of modules. To avoid the dead
spaces in a multi-module design, there is just a single large EM module
and one inner hadronic (IH) module.
Outside the EM and IH, there are concentric rings
of 16 middle and outer hadronic modules (MH and OH).
The azimuthal boundaries of
the MH and OH modules are also offset to prevent cracks through which particles
could penetrate the calorimeter. This makes the D{\O} detector almost
completely hermetic and provides an accurate measurement of missing
transverse energy. Due to increase in background and loss of
tracking efficiency for $|\eta|>2.5$, 
electron and photon candidates
are restricted to $1.5<|\eta|<2.5$ in the EC.

	In the transition region between the CC and EC
($0.8 \leq | \eta | \leq 1.4$),
there is a large amount of uninstrumented material in the form of cryostat
walls, stiffening rings, and module endplates.
To correct for energy deposited in the
uninstrumented material, we use two segmented
(0.1~$\times$~0.1 in $\eta\times\phi$)
arrays of scintillation counters,
called intercryostat detectors.  In addition,
separate single-cell structures called ``massless gaps" are mounted on
the end plates of the CC-FH modules and on the front plates of EC-MH and EC-OH
modules, and are used to correct showers in this region of the detector.

	The Main Ring beam pipe passes
through the outer layers of the CC, ECN and ECS. Beam losses
from the Main-Ring cause
energy deposition in the calorimeter that can bias the energy measurement.
The data acquisition system either stops recording data during periods of
Main-Ring activity near the D{\O} detector, or flags such events. 

\subsection{Muon Detectors}
\label{muon-detectors}

The D{\O} muon detector is designed to identify muons and
to determine their trajectories and momenta.
It is located outside of the calorimeter, and is divided in two
subsystems: the Wide Angle Muon Spectrometer and the Small Angle
Muon Spectrometer. Since the calorimeter
is thick enough to absorb most of the debris from electromagnetic
and hadronic
showers, muons can be identified with great confidence. The
muon system is not used in this analysis, and is therefore not discussed
any further.

\section{Multilevel Trigger and Data Acquisition Systems}
\label{trigger-daq}

The D{\O} trigger system is
a multilayer hierarchical system. Increasingly complex tests are
applied to the data at each successive stage to reduce background.

        The first stage, called Level 0 (L0), consists of
two scintillator arrays mounted on the front surfaces of the EC
cryostats, perpendicular to the beam direction. Each array covers
a partial region of pseudorapidity for 1.9~$< |\eta| <$~4.3,
with nearly complete
coverage over the range 2.2~$< |\eta| <$~3.9.
The L0 system is used to detect the
occurrence of an inelastic $p{\bar p}$ collision, and serves as the luminosity
monitor for the experiment.  In addition, it provides fast information on
the $z$-coordinate of the primary collision vertex, by measuring the
difference in arrival time between particles hitting the north and south
L0 arrays; this is used in making preliminary
trigger decisions.  A slower, more accurate measurement of the
position of the interaction vertex, and an indication of the possible
occurrence of multiple
interactions, are also made available for subsequent trigger decisions.
The L0 trigger is $\approx$
99\% efficient for non-diffractive inelastic collisions. The output rate
from L0 is on the order of 150 kHz at a typical luminosity of
$1.6\times 10^{31}{\rm cm}^{-2}$ ${\rm s}^{-1}$. 

       The next stage of the trigger is called Level 1 (L1). It 
combines the results from individual L1 components into a
set of global decisions that command the readout of the digitization
crates. It also interacts with the Level 2 trigger (L2).
Most of the L1 components, such as the
calorimeter triggers and the muon triggers, operate within the 3.5 $\mu s$
interval between beam crossings, so that all events are examined.
However, other components, such as the TRD trigger and several components
of the calorimeter and muon triggers, called Level 1.5 trigger (L1.5),
can require more time.
The goal of the L1 trigger is to reduce the event rate 
to 100--200 Hz. The primary input for the L1
trigger consists of 256 trigger terms, each of which
corresponds to a single bit,
indicating that some specific requirement is met. These 256 terms are
reduced to a set of 32 L1 trigger bits by a
two-dimensional AND-OR logic network. An event is said to pass L1 if at least
one of these 32 bits is set.  The L1 trigger also uses information based on
Main Ring activity. To
prevent saturation of the trigger system by processes with large cross
sections, such as QCD multijet production, any particular contributor to
the L1 trigger can be prescaled.

        The L1 calorimeter trigger covers the region up to $|\eta| <$~4.0
in trigger
towers of 0.2~$\times$~0.2 in $\eta-\phi$ space. These towers are
subdivided longitudinally into electromagnetic and hadronic trigger
sectors. The output of the L1 calorimeter trigger corresponds to
the transverse energy deposited in these sectors and towers.

	For the 1993--1995 collider run, an L1.5 trigger for the calorimeter was
implemented using the L1 calorimeter trigger data and filters
based on neighbor sums and ratios of the EM and total transverse
energies.

        When an event satisfies the L1 trigger, the data are passed on the
D{\O} data acquisition pathways to a farm of 48 parallel
microprocessors, which serve as event builders as well as the L2 trigger
system. The L2 system collects the digitized data
from all elements of the detector and trigger blocks for events that
successfully pass Level 1.
It applies sophisticated algorithms to the data
to reduce the event rate to about 2 Hz before passing the accepted events on
to the host computer for monitoring and recording. The data for
a specific event are sent over parallel paths to memory modules in
specific selected nodes. The accepted data are collected and
formatted in final form in the nodes, and the L2 filter algorithms are
then executed. 

        The L2 filtering process in  each node is built around a series of
filter tools. Each tool has a specific function related to the identification
of a type of particle or event characteristic. There are tools to
recognize jets, muons, calorimeter EM clusters, tracks associated with
calorimeter clusters, $\sum E_T$ (sum of transverse energies of jets),
and $\MET$ (imbalance in transverse energy). Other tools recognize
specific noise or background conditions.  There are 128
L2 filters available. If all of the L2 requirements (for at least one
of these 128 filters) are satisfied, the event is said to pass L2 and it
is temporarily stored on disk before being transferred to an 8 mm magnetic
tape.

Once an event is passed by an L2 node, it is transmitted to the host
cluster, where it is received by the data logger, a program running on one
of the host computers. This program and others associated with it are
responsible for receiving data from the L2 system and copying it to
magnetic tape, while performing all necessary bookkeeping tasks (e.g., time
stamping, recording the run number, an event number, etc.).
Part of the data is sent to an event pool for online monitoring.

\subsection{Electron Trigger}
\label{electron-trigger}

To trigger on electrons, L1 requires the transverse energy in the EM
section of a trigger tower to be above a programmable threshold. 
The L2 electron algorithm then uses the full segmentation of the EM calorimeter
to identify electron showers. Using the trigger towers that are above
threshold at L1 as seeds, the algorithm forms clusters that include
all cells in the four EM layers and the first FH layer in a region of
$\Delta\eta\times\Delta\phi$ = 0.3~$\times$~0.3, centered on the tower
with the highest $E_T$. The longitudinal and transverse energy profile of
the cluster must satisfy the following requirements:
(i) the fraction of the cluster energy in the EM section (the EM fraction)
must be above
a threshold, which depends on energy and detector position; and
(ii) the difference between the energy
depositions in two regions of the third EM layer,
covering $\Delta\eta\times\Delta\phi$ =
0.25~$\times$~0.25 and 0.15~$\times$~0.15, and centered on the cell with
the highest $E_T$, must be within a window that depends on the
total cluster energy.

\subsection{Jet Trigger}
\label{jet-trigger}

The L1 jet triggers require the sum of the transverse energy in the EM and
FH sections of a trigger tower ($\Delta\eta\times\Delta\phi=0.2\times 0.2$)
to be above a programmable threshold. 
The L2 jet algorithm begins with an $E_T$-ordered list of towers that are
above threshold at L1. At L2, a jet is formed by placing a cone of given
radius ${\cal R}$, where ${\cal R} = \sqrt{\Delta\eta^2 + \Delta\phi^2}$,
around the
seed tower from L1. If another seed tower lies within the jet cone, it
is passed over and not allowed to seed a new jet. The summed $E_T$ in all
of the towers included in the jet cone defines the jet $E_T$. If any two
jets overlap, then the towers in the overlap region are added into the jet
candidate that is formed first. To filter out events, requirements on
quantities such as the minimum transverse energy of a
jet, the minimum transverse size of a jet, the minimum number of jets,
and the pseudorapidity of jets, can be imposed at this point. 

\subsection{Missing Transverse Energy Trigger}
\label{met-trigger}

        Rare and interesting physics processes often involve production of
weak\-ly interacting particles such as neutrinos. These particles usually
can not
be detected directly. However, assuming
momentum conservation in a collision allows
the momenta of such particles to be inferred from the vector sum of the
momenta of the observed particles. Since the energy flow
near the beamline is largely undetected, such calculations are realistic only
in the plane transverse to the beam. 
The negative of the vector sum of the momenta of the detected particles 
is referred to as missing $E_T$ and denoted by $\MET$; it is used
as an indicator of the presence of weakly interacting particles. 
At L2, $\MET$ is computed using the vector sum of all calorimeter and
intercryostat detector
cell energies with respect to the $z$ position of the interaction vertex, which is
determined from the timing of the hits in the L0 counters.

\section{Particle Identification}
\label{particle-id}

\subsection{Electron}
\label{electron-id}

Electrons and photons are identified by the properties of the shower in the
calorimeter.
The algorithm loops over all EM towers
($\Delta\eta\times\Delta\phi=0.1\times 0.1$)
with energy $E>50$ MeV, and connects the
neighboring tower with the next highest energy. The cluster energy is then
defined as the sum of the energies of the EM towers and the energies in the
corresponding first FH layer. The ratio of the energy in the EM cluster to
the total energy (EM energy summed with the corresponding
hadronic layers), defined as the EM fraction, is used to discriminate electrons
and photons from hadronic showers.
A cluster must
pass the following criteria to be an electron/photon
candidate: (i) the EM fraction must be greater than 90\%
and (ii) at least 40\% of the energy must be contained in a single
$0.1\times 0.1$
tower. To distinguish electrons from photons, we search for a track
in the central detector that extrapolates to the EM
cluster from the primary interaction vertex 
within a window of $|\Delta\eta| \le 0.1$, and $|\Delta\phi| \le 0.1$.
If one or more tracks are found, the object is classified as an electron
candidate.
Otherwise, it is classified as a photon candidate.

\subsubsection{Selection Requirements}
\label{electron-selection}

        The spatial development of EM showers is quite different 
from that of hadronic showers and
the shower shape information can be used to differentiate
electrons and photons from hadrons. The following variables are used for
final electron selection:

\noindent (i) Electromagnetic energy fraction. This quantity
is based on the observation that electrons deposit almost
all of their energy in the EM section of the calorimeter,
while hadron jets are far more penetrating (typically only 10\% of their
energy is deposited in the EM section of the calorimeter). It is defined
as the ratio of EM energy to the total shower energy.
Electrons are required to have at least
95\% of their total energy in the EM calorimeter. This requirement loses
only about 1\% of all electrons.

\noindent (ii) Covariance matrix ($H$-matrix) $\chi^2$. The shape of
any shower
can be characterized by the fraction of the cluster energy deposited
in each layer and tower of the calorimeter. These fractions are correlated,
i.e., 
an electron shower deposits energies according to the expected
transverse and longitudinal shapes of an EM shower and a hadron shower
following the typical development of a hadronic shower.
To obtain good discrimination
against hadrons, we use a covariance matrix technique.
The observables in this method are the fractional energies in Layers
1, 2, and 4 of the EM sector
and the fractional energy in each cell of a $6\times 6$
array of cells in Layer 3 centered on the most energetic tower in the EM
cluster. To take account of the dependence of the shower shape on
energy and on the position of the primary interaction vertex,
we use the logarithm of the shower energy and the
z-position of the event vertex as the remaining input observables.
The event vertex is determined by extrapolating CDC
tracks to the $z$ axis, and for more than one possibility,
the vertex associated with the highest number of tracks
is chosen as the event vertex.
Using these 41 variables, covariance matrices are constructed
for each of the 37 detector
towers (at different values of $\eta$) based on Monte Carlo generated electrons.
The Monte Carlo showers are tuned to make them agree with our test
beam measurements of the shower shapes.
The 41 observables for any given shower can be compared with the parameters of
the appropriate covariance matrix to define a $\chi^2$, which is to be
be less than 100 for electron candidates in the CC and less than 200 for the
EC. This requirement loses about 5\% of all true electrons.

\noindent (iii) Isolation. The decay electron from a $W$ boson should not be
close to any other object in the event. This is quantified by the
isolation fraction.  If $E(0.4)$ is the energy deposited in all calorimeter
cells within the cone ${\cal R} < 0.4$ around the direction of the electron, and
$EM(0.2)$ is the energy deposited in only the EM calorimeter in the cone 
${\cal R}< 0.2$, the isolation variable is then defined as the ratio
${\cal I} = [E(0.4)-EM(0.2)]/EM(0.2).$
The requirement ${\cal I} < 0.1$ loses only 3\% of the electrons from $W$
boson decays.

\noindent (iv) Track-match significance. An important source of
background for electrons is the photon from the decay of $\pi^0$ or $\eta$
mesons. Such photons do not produce tracks, but their trajectories can overlap
with those of nearby charged particles, thereby simulating electrons.
This background can
be reduced by demanding a good spatial match between the energy
cluster in the calorimeter and nearby charged tracks.
The significance $S$ of the mismatch between
these quantities is given by
$ S= [({\Delta\phi}/{\delta_{\Delta\phi}})^2 +
({\Delta z}/{\delta_{\Delta z}})^2]^{1/2}$,
where $\Delta\phi$ is the azimuthal mismatch, $\Delta z$ the
mismatch along the beam axis, and the $\delta$ are the resolutions of these
variables. This form for $S$ is appropriate for the central calorimeter.  For
the end calorimeter, $r$ replaces $z$. Requiring $S<5$ accepts 95(78)\%
of the CC(EC) electrons reconstructed in the central tracker.

\noindent (v) Track-in-road. All electrons from
$W\rightarrow e\nu$ decays are required to have a partially
reconstructed track along the trajectory between the energy cluster 
in the calorimeter and the interaction vertex.  
This requirement is found to reject 16(14)\% of
CC(EC) electrons from $W$ boson decay.

        In our analysis, we combine the above quantities to form the
electron identification criteria. A summary of the selection requirements
and their acceptance
efficiencies is listed in Table~\ref{table:eid}
(See Sect\ref{detection-efficiency}).

\begin{table}[htb]
\caption{Electron selection requirements and their acceptance efficiencies
for $W\rightarrow e\nu$ events.}
\label{table:eid}
\begin{center}
\begin{tabular}{c|lclc}
Selection &\multicolumn{2}{c}{CC} & \multicolumn{2}{c}{EC} \\
requirement &  & $\varepsilon$ & & $\varepsilon$ \\ \hline\hline
$H$-matrix $\chi^2$ & $<100$  & 0.946$\pm$0.005 & $<200$  & 0.950$\pm$0.008
\\ \hline
EM fraction       & $>0.95$ & 0.991$\pm$0.003 & $>0.95$ & 0.987$\pm$0.006
\\ \hline
Isolation         & $<0.10$ & 0.970$\pm$0.004 & $<0.10$ & 0.976$\pm$0.007
\\ \hline
Track match       & $< 5  $ & 0.948$\pm$0.005 & $< 5  $ & 0.776$\pm$0.012
\\ \hline
Track-in-road     &         & 0.835$\pm$0.009 &         & 0.858$\pm$0.006
\end{tabular}
\end{center}
\end{table}

\subsubsection{Electromagnetic Energy Corrections}
\label{em-energy-corrections}

The energy scales of the calorimeters were originally set through 
calibration in a test-beam.
However, due to differences in conditions between the
test beam and the D{\O} environment, additional corrections had to be
implemented.

The EM energy scales for the calorimeters were determined by comparing
the measured masses of $\pi^0\rightarrow \gamma\gamma$,
$J/\psi\rightarrow ee$, and $Z \rightarrow ee$ to their
known values.
If the electron energy measured in the
calorimeter and the true energy are related by $E_{\rm meas} 
= \alpha E_{\rm true} + \delta$,
the measured and true mass values are, to first order, related
by $m_{\rm meas} = \alpha m_{\rm true} + \delta f$,
where the calculable variable $f$ reflects
the topology of the decay.
To determine $\alpha$ and $\delta$, we fit the Monte Carlo prediction
to the observed resonances, with $\alpha$ and $\delta$ as free
parameters~\cite{D0Wmass}.
The values of $\alpha$ and $\delta$ are found
to be $\alpha=0.9533\pm0.0008$ and $\delta=-0.16^{+0.03}_{-0.21}$ GeV for
the CC and $\alpha=0.952\pm0.002$ and $\delta=-0.1\pm0.7$ GeV for the EC.

\subsubsection{Energy Resolution}
\label{em-energy-resolution}

        The relative energy resolution for electrons and photons in the CC 
is expressed by the empirical relation
$\left(\frac{\sigma}{E}\right)^2 = C^2 + \frac{S^2}{E_T} +
\frac{N^2}{E^2}$,
where $E$ and $E_T$ are the energy and transverse energy of 
the incident electron/photon, $C$ is a
constant term from uncertainties in calibration, $S$ reflects
the sampling fluctuation
of the liquid argon calorimeter, and $N$ corresponds to a contribution from
noise. 
For the EC, the $E_T$ in the relation is replaced by $E$.
The sampling and noise terms are based on results from the test beam.
The noise term measured at the test beam agrees with the one
obtained in the collider environment (based on the width of pedestal
distributions). The constant term is tuned to match the mass resolution
of both observed and simulated $Z\rightarrow ee$ events.
Table~\ref{table:e-res} lists these parameters.

\begin{table}[htb]
\caption{Parameters for describing the energy resolution of electrons and
photons.}
\begin{center}
\begin{tabular}{ccc}
Quantity &  CC   & EC     \\ \hline
$C$      & 0.017 & 0.009 \\
$S$ ($\sqrt{\rm GeV}$)      & 0.14 & 0.157 \\
$N$ (GeV)      & 0.49 & 1.140 \\
\end{tabular}
\label{table:e-res}
\end{center}
\end{table}

\subsection{Jets}
\label{jet-id}

In our analysis, jets are reconstructed using a fixed-cone
algorithm with radius ${\cal R} =
\sqrt{\Delta\eta^2 + \Delta\phi^2}=0.5$.
The algorithm forms preclusters of contiguous cells using a 
radius of ${\cal R}_{\rm precluster} = 0.3$ centered on the tower with
highest $E_T$. Only towers with $E_T>$~1 GeV are
included in preclusters. These preclusters serve as the starting points for
jet reconstruction.
An $E_T$-weighted center of gravity is then formed using the $E_T$ of all towers
within a radius ${\cal R}$ of the center of the cluster,
and the process is repeated until the
jet becomes stable. A jet must have $E_T>$~8 GeV. If two jets share energy,
they are combined or split, based on the fraction of overlapping energy
relative to the $E_T$ of the lesser jet. If this shared fraction exceeds
50\%, the jets are combined.

Although the
${\cal R} = 0.3$ cone algorithm is more efficient for jet finding
than our larger
cone size, which leads to undesired merging of jets for high-$p_T$ $W$ or
$Z$ bosons, the relatively large uncertainties in the
measurement of jet-energy for the ${\cal R} = 0.3$ cones
negate their advantage, and
we therefore choose to use
the ${\cal R} = 0.5$ cone algorithm for
our studies.

\subsubsection{Selection Requirements}
\label{jet-selection}

        To remove jets produced by cosmic rays,
calorimeter noise, and interactions in the Main Ring, 
we developed a set of requirements
based on Monte Carlo studies of jets in such environments 
and on data on noise taken with and without colliding beams.
The variables used are:

\noindent (i) Electromagnetic energy fraction ($emf$). As for
electrons, this quantity is defined as the fraction of the total
energy deposited
in the electromagnetic section of the calorimeter.
A requirement on this quantity removes electrons, photons
and false jets from the jet sample.
Electrons and photons typically have a high EM fraction.
False jets are caused mainly by background from the
Main Ring or by noisy or ``hot" cells, and therefore generally
do not contain energy in the EM section, thereby yielding very low EM
fractions.
Jets with $0.05<emf<0.95$ are defined as acceptable in this analysis.
The efficiency of this requirement is 99.9\% at $E_T=20$ GeV and decreases to
99.6\% at 100 GeV. 
   
\noindent (ii) Hot cell energy fraction ($hcf$). The $hcf$ is defined
as the ratio of the energy in the cell of second highest $E_T$
to that of the cell with highest $E_T$ within a jet.
A requirement on this quantity is imposed to remove
events with a large amount of noise in the
calorimeter.
Hot cells can appear when a discharge occurs between electrodes within a
cell; often this does not affect neighboring cells.
In this case, $hcf$ is small, which signals a problem, since
the $hcf$ for a jet should not be small 
because the energy is expected to be distributed over
cells. If most of the energy is concentrated in a single cell,
it is very
likely to be a false jet reconstructed from discharge noise.  For good
jets, $hcf$ is found to be greater than 0.1.
The efficiency of this requirement is 97.3\% at
$E_T=20$ GeV and decreases to 96.9\% at 100 GeV.
   
\noindent (iii) Coarse hadronic energy fraction ($chf$). This quantity is
defined as the fraction of jet energy deposited in the coarse hadronic
section of the calorimeter.
The Main Ring at D\O\ passes through the
CH modules, and any energy deposition related to the Main Ring
will be concentrated in
this section of the calorimeter. Such jets
tend to have more than 40\% of their energy in the CH
region, while standard jets have less than 10\% of their energy
in this section of the calorimeter. All acceptable jets are
therefore required to have $chf<0.4$.
The efficiency of this requirement is 99.6\% at $E_T=20$ GeV
and decreases to 99.3\% at 100 GeV.

\subsubsection{Hadronic Energy Corrections}
\label{had-energy-corrections}

Since the measured jet energy is usually not equal to the energy of the
original parton that formed the jet, corrections are needed to minimize
any systematic bias.
Jet energy response affected by non-uniformities in the calorimeter,
non-linearities in the response to hadrons,
emission of particles outside of the ${\cal R}=0.5$ cone (often referred
to as 
out-of-cone showering), noise due to the radioactivity of uranium, 
and energy overlap from the products of soft
interactions of spectator partons within the proton and the antiproton
(``underlying event").
The first two effects are estimated using a method called
Missing-$E_T$ Projection Fraction (MPF)~\cite{Ehad_scale}.

        The MPF method is based on events that contain a single isolated
EM cluster (due to a photon or a jet that fragmented mostly into neutral
mesons), and one hadronic jet located opposite in $\phi$,
and no other objects in the event.
It is assumed that such events do not have energetic neutrinos
so that any missing transverse energy
can be attributed to a mismeasurement of the hadronic jet.
The EM-cluster energy is corrected using the electromagnetic
energy corrections described above.
Projecting the corrected $\MET$ along the jet axis determines
corrections to the jet energy.
This correction is averaged over many events in the sample to
obtain a correction as a function of jet $E_T$, $\eta$, and
electromagnetic content of the jet. 
The hadronic energy correction is 20\% at $E=20$ GeV and 15\% at $E=100$ GeV,
and gradually approaches 10\% at high $E$.

The impact of out-of-cone showering is estimated using Monte Carlo jet
events.
Effects due to the underlying events and uranium noise are
determined in separate studies using minimum-bias event data.
(Minimum-bias data corresponds to inclusive inelastic
collisions collected using only the L0 trigger.)

\subsubsection{Energy Resolution}
\label{had-energy-resolution}

The jet energy resolution has been studied by examining
momentum balance in dijet events~\cite{E_resolution}.
The formula used for parametrizing
the relative jet energy
resolution is
$\left(\frac{\sigma}{E}\right)^2 = C^2 + \frac{S^2}{E} +
\frac{N^2}{E^2}$.
Table \ref{table:jetres} shows the values of the parameters for different
$\eta$ regions of the calorimeter.

\begin{table}[htb]
\caption{Jet energy resolution for different regions of the calorimeter.}
\begin{center}
\begin{tabular}{c  r@{$\pm$}l  r@{$\pm$}l  r@{$\pm$}l}
$\eta$ Region  & \multicolumn{2}{c}{$C$}& \multicolumn{2}{c}
{$S$ ($\sqrt{\rm GeV}$)} & \multicolumn{2}{c}{$N$ (GeV)}           \\
\hline
$|\eta| <$ 0.5       & 0.00 & 0.01 & \hspace*{12pt}0.81 & 0.02 & 7.07 & 0.09 \\
0.5 $< |\eta| <$ 1.0 & 0.00 & 0.01 & 0.91 & 0.02 & \hspace*{12pt}6.92 & 0.09 \\
1.0 $< |\eta| <$ 1.5 & 0.05 & 0.01 & 1.45 & 0.02 & 0.00 & 1.40 \\
1.5 $< |\eta| <$ 2.0 & 0.00 & 0.01 & 0.48 & 0.07 & 8.15 & 0.21 \\
2.0 $< |\eta| <$ 3.0 & 0.01 & 0.58 & 1.64 & 0.13 & 3.15 & 2.50 \\
\end{tabular}
\label{table:jetres}
\end{center}
\end{table}

\subsection{Neutrinos: Missing Transverse Energy}
\label{netrinos}

        The presence of neutrinos in an event
is inferred from the \MET. In this analysis we
assume that the $\MET$ in each candidate event corresponds to the
neutrino from the decay $W\rightarrow e\nu$.

\subsubsection{Missing $E_T$}
\label{met}

        The missing transverse energy in the calorimeter
is defined as
$\MET = (\MET_{x}^2 + \MET_{y}^2)^{1/2}$,
where
$\MET_{x} = -\sum_{i} E_i \sin(\theta_i)\cos(\phi_i) -
\sum_{j}\Delta E_x^j$ and 
$\MET_{y} = -\sum_{i} E_i \sin(\theta_i)\sin(\phi_i) -
\sum_{j}\Delta E_y^j$.
The first sum (over $i$) is over all cells in the calorimeters, intercryostat
detectors
and massless gaps (see Sec.~\ref{calorimeters}).
The second sum (over $j$) is over the $E_T$ corrections
applied to all electrons and jets in the event. 
This can be used to
estimate the transverse momentum of any neutrinos in an event that does not
contain muons, which deposit only a small
portion of their energy in the calorimeter. The total missing
$E_T$ is missing $E_T$ from the calorimeter corrected for
the transverse momenta of any observed
muon tracks. Since this analysis does not use muons,
we will refer to the $\MET$ based on the calorimeters as the true $\MET$.

\subsubsection{Resolution in $\MET$}
\label{met-resolution}

        For an ideal calorimeter, the magnitude of the components of
the $\MET$ vector would sum to zero for events with no true source of $\MET$.
However, detector noise and energy resolution in the measurement of jets,
photons, and electrons contribute to the $\MET$. In
addition, non-uniform response in the detector
also results in $\MET$.
The $\MET$ resolution for our candidate events is parameterized as
$\sigma = 1.08{\rm GeV} + 0.019 (\sum E_T)$,
and is based on studies of minimum-bias data~\cite{E_resolution}. 
The $\sum E_T$
used in the parameterization is quite reasonable because the greater the
total amount of transverse energy in the event, the larger the possibility
for its mismeasurement.

\section{Data Sample}
\label{data-sample}

The analysis of the $WW/WZ\rightarrow e\nu jj$ process is based on
data taken during the 1993--1995 Tevatron
Collider run (called Run 1b).
The L0 trigger is used to check the presence of
an inelastic collision, but is not included in
the trigger conditions for $W$-boson data.
This was done to allow studies of diffractive $W$-boson production. 
Our analysis uses the collected $W\rightarrow e\nu$ data sample, with the
L0 trigger requirement imposed offline.
The L1 trigger used in this analysis
(called the {\tt EM1\_1\_HIGH} trigger) 
requires the presence of an electromagnetic trigger tower with 
$E_T>10$ GeV.
The L1.5 trigger then requires the L1 trigger tower to have $E_T>15$ GeV and
checks that the electromagnetic
fraction is greater than 85\%.
The L2 component of the trigger
(called the {\tt EM1\_EISTRKCC\_MS} trigger) 
requires an isolated electron candidate with $E_T>20$ GeV
that has a shower shape consistent with that of an electron
and $\MET>15$ GeV.

Additional conditions are imposed on the data to further
reduce background.
Triggers that occur at the times when a proton bunch in the Main Ring passes
through the detector are not used in this analysis.
Similarly, triggers that occur during the first 0.4 seconds of the
2.4-second antiproton production cycle are rejected.
Data taken during periods when
the data acquisition system or the detector sub-systems
malfunctioned are also discarded.
With these trigger requirements,
the integrated luminosity of the data sample 
is estimated to be
$82.3\pm4.4$ ${\rm pb}^{-1}$~\cite{luminosity}.
The efficiency and turn-on of the L2 trigger
are described in Ref.~\cite{Kelly}.
The trigger efficiency for signal is ($98.1\pm 1.9$)\%. 

Data samples that satisfy two other L2 triggers,
the {\tt EM1\_ELE\_MON} and {\tt ELE\_1\_MON} triggers,
are used for background studies.
These triggers select events that have an electron candidate with
$E_T>20$ GeV and $E_T>16$ GeV, respectively.
The electron candidates in these samples must pass the standard
shower-shape requirements, but not the isolation requirement.
These triggers use the same L1 and L1.5 conditions as the
trigger used for signal.

\section{Event Selection}
\label{event-selection}

$WW/WZ\to e\nu jj$ candidates are selected by searching for events
with an isolated high-$E_T$ electron, large $\MET$,
and at least two high-$E_T$ jets.
Electrons in the candidate sample must be in the $|\eta|<1.1$ region
but away from the boundaries between
calorimeter modules in $\phi$ ($\Delta\phi>0.01$), or
within the region $1.5<|\eta|<2.5$.
Jets in the candidate sample must be in the region $|\eta|<2.5$.

The $W\rightarrow e\nu$ decay is defined through the presence of
only a single isolated
electron with $E_T^e>25$ GeV and $\MET>25$ GeV in the event.
The transverse mass of the electron and neutrino  ($\MET$)
system is required to be
$M_T > 40$~GeV/$c^2$, where 
$M_T= \{2 E_T^e \MET [1-\cos(\phi_{e}-\phi_{\nu})]\}^{1/2}$. 
The requirement on the electron $E_T$
is sufficiently high to provide an efficiency that is independent of $E_T$
(the hardware threshold of $20$ GeV).
Requiring only one electron reduces background from
$Z\rightarrow ee$ production.
The requirements on $\MET$ and $M_T$ reduce the background
contribution from misidentified electrons.

The $W/Z \rightarrow jj$ decay is defined by
requiring at least two jets with $E^j_T > 20$ GeV and
an invariant mass of the two-jet system consistent 
with that of the $W$
or $Z$ boson ($50 < M_{jj} < 110$~GeV/$c^2$). 
The dijet invariant mass
($M_{jj}$) is calculated via 
$M_{jj} = \{2 E_T^{j1} E_T^{j2} 
[\cosh(\eta_{j1} - \eta_{j2}) -\cos(\phi_{j1}-\phi_{j2})]\}^{1/2}$.
If there are more than two jets in the event, the two jets with the highest
dijet invariant mass are chosen to represent the $W$ (or $Z$) decay. 

The difference between the $p_T$ values of the $e\nu$ and the two-jet systems is
used to reduce backgrounds.
For $WW$ or $WZ$ production, the $p_T(e\nu) - p_T(jj)$
distribution should be peaked near zero and have a symmetric Gaussian shape,
with the width of the Gaussian distribution determined primarily by the jet
energy resolution. On the other hand, for background such as
$t\bar{t}$ production (see Sec.~\ref{background}), the
distribution should be broader and
asymmetric (shifted to positive values)
due to additional $b$-quark jets in the events. Our analysis therefore requires
$|p_T(e\nu)-p_T(jj)| < 40$ GeV/$c$.

\begin{figure}[hbt]
\epsfxsize=3.4in
\centerline{\epsffile{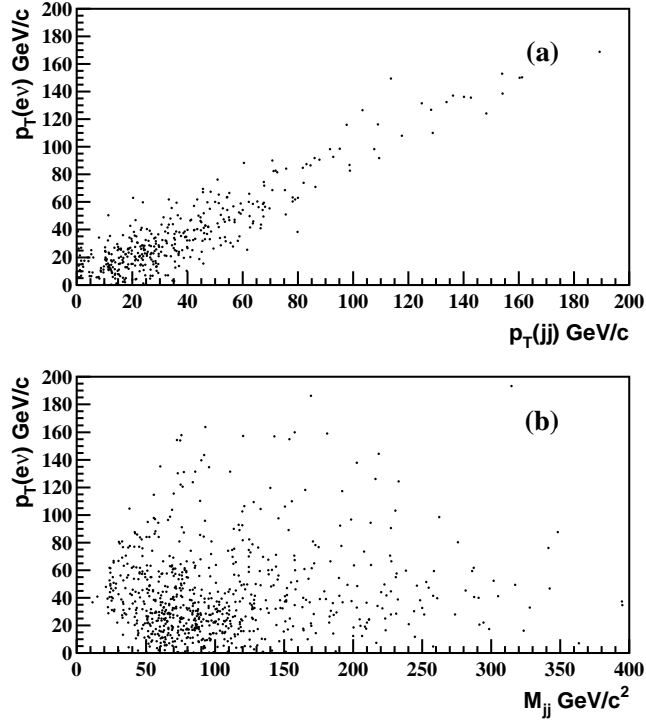}}  
\caption{Scatter plots of (a) $p_T(e\nu)$ vs $p_T(jj)$ and (b)
$p_T(e\nu)$ vs $M_{jj}$.}
\label{fig:pt_mjj}
\end{figure}

The data satisfying the above selection criteria yield 399 events. 
Figure~\ref{fig:pt_mjj}a  shows a scatter plot of $p_T(e\nu)$
vs $p_T(jj)$ for candidate events that satisfy the two-jet mass
requirement. The width
of the band reflects both the resolution and the true spread in the $p_T$
values. Figure~\ref{fig:pt_mjj}b shows a scatter plot of $p_T(e\nu)$ vs
$M_{jj}$ without the imposition of the two-jet mass requirement.

\section{Detection Efficiency}
\label{detection-efficiency}
\subsection{Electron Selection Efficiency}
\label{electron-selection-efficiency}

The efficiency of electron selection is
studied using the $Z\rightarrow ee$ event
sample from the 1993--1995 Tevatron collider run
using the {\tt EM2\_EIS\_HI} trigger.
$Z\rightarrow ee$ events were selected at L1 and
L1.5 by requiring two EM towers with $E_T > 7$ GeV at 
L1, and at least one tower with $E_T > 12$ GeV with more than
$85$\% of its energy in
the EM section of the calorimeter. At L2, the trigger required two 
electron candidates with $E_T > 20$ GeV that satisfied electron
shower-shape and isolation requirements.
To select an
unbiased sample of electrons, we use
events in which one of the electrons passes the tag quality requirements:
EM fraction $>0.90$, Isolation $< 0.15$, $H$-matrix
$\chi^2 <$ 100(200) for CC(EC), and track-match significance $< 10$.
The second electron in the event is then assumed to be unbiased.
If both electrons pass the tag requirements,
the event contributes twice to the sample.
The efficiency of a selection requirement for electrons is given by
$\varepsilon = (\varepsilon_s - \varepsilon_b f_b)/(1 - f_b)$,
where $\varepsilon_s$ is the efficiency measured in the signal
region, $\varepsilon_b$ is the efficiency measured in the background
region, and $f_b$ is the ratio of the number of background events in the
signal region to the total number of events in the signal region.  The
signal region is defined as the region of the $Z$ mass peak
($86<m_{ee}<96$ GeV/$c^2$), and
the background regions are defined as $61<m_{ee}<71$
GeV/$c^2$ and $111<m_{ee}<121$ GeV/c$^2$. We determine $f_b$ in the region
of the signal
using an average of the number of events in the background
regions. 
The systematic uncertainties on the efficiencies
are estimated from a comparison with efficiencies obtained using an
alternative method that fits the invariant mass spectrum of two electrons
to the sum of a Breit-Wigner form
convoluted with a Gaussian and a linear dependence for the background.
Efficiencies from
the two methods agree within their uncertainties.
The track-in-road efficiency is
estimated in a similar manner,
except that EM clusters with no matching track are included as
unbiased electrons in the sample.
Table~\ref{table:eid}
summarizes the electron efficiencies.
Although these efficiencies are based mainly on $Z$ events with few jets,
the corrections for $\ge2$ jets are small.

\subsection{$W/Z\rightarrow jj$ Selection Efficiency}
\label{wjj-selection-efficiency}

\begin{figure}[htb]
\epsfxsize=3.4in
\centerline{\epsffile{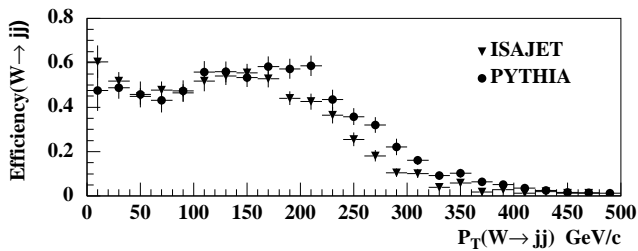}}  
\caption{Efficiency for $W\rightarrow jj$ selection as a function of
$p^W_T$.
The decrease in the efficiency at high $p_T$ is due to the merging of the
two jets from the decay of a $W$ boson.
}
\label{fig:wjj_eff}
\end{figure}

The $W/Z\rightarrow jj$ selection efficiency is estimated
using Monte Carlo $WW/WZ\rightarrow e\nu jj$ 
events generated with the {\sc isajet}~\cite{isajet}
and {\sc pythia}~\cite{pythia}
programs, followed by a detailed simulation of the D{\O} detector, and
parametrized as a function of $p^W_T$.
Figure~\ref{fig:wjj_eff} shows the $W/Z\rightarrow jj$ detection efficiency
$\epsilon(W\rightarrow jj)$ calculated as the ratio of
events after the imposition of the two-jet selection requirements
relative to the initial
number of events.
At low $p_T$, the detection efficiency is artificially elevated due to
the presence of additional jets
from initial- and final-state gluon radiation (ISR/FSR) that are
mislabeled as being decays of $W$ or $Z$ bosons.
The decrease in the efficiency at high $p_T$ is due to the merging of the
two jets from a $W$ or a $Z$ boson.  
The results obtained from {\sc isajet} are used to
estimate the efficiencies for identifying the $WW/WZ$ process.

The estimated $W/Z\rightarrow jj$ efficiency is affected by the
jet energy scale, the accuracy of the 
ISR/FSR simulation, the accuracy of the parton
fragmentation mechanism, and the statistics of the Monte Carlo samples.

The energy-scale correction has an
uncertainty that decreases from 5\% at jet $E_T=20$ GeV
to 2\% at 80 GeV, and then
increases to 5\% at 350 GeV.
The effect of this uncertainty has been
studied by recalculating the efficiency with the jet energy scale changed by
one standard deviation.
The largest relative change in the accepted number of
events is found to be 3\%.

To estimate the uncertainty due to the accuracy of the ISR/FSR
simulation and of the parton fragmentation mechanism, we use
the $W/Z\rightarrow jj$
efficiency based on Monte Carlo samples generated with {\sc pythia}.
The efficiency
obtained using {\sc isajet} is lower than that for {\sc pythia}, but by
less than 10\%. We use the efficiencies
from {\sc isajet} because they provide smaller yields of
$WW/WZ$ events and therefore weaker limits on anomalous
couplings.
We define one-half of the largest difference in {\sc isajet/pythia}
efficiency estimations (5\%) as the systematic uncertainty attributable
to the choice of event generator.

\subsection{Overall Selection Efficiency}
\label{overall-selection-efficiency}

The overall detection efficiency for $WW/WZ\rightarrow e\nu jj$ events
assuming SM couplings is calculated using
two MC methods, coupled with electron-selection
and trigger efficiencies measured from data.
The first MC method uses the {\sc isajet} event generator followed
by a detailed simulation of the D{\O} detector.
The second MC method uses the event
generator of Ref.~\cite{Hagiwara90} and a fast simulation program
to characterize the response of the detector.  
{\sc isajet} used the {\small CTEQ2L}~\cite{cteq2l} parton
distribution functions to
simulate 2500 $WW\rightarrow e\nu jj$ events
and 1000 $WZ\rightarrow e\nu jj$
events with SM couplings. The event selection efficiency for
for the $WW\rightarrow e\nu jj$ signal
is estimated as $\epsilon_{WW}=(13.4\pm0.8)$\%,
and $\epsilon_{WZ}=(15.7\pm1.4)$\% for the $WZ\rightarrow e\nu jj$ signal,
where the errors are statistical. The combined efficiency
for $WW/WZ\rightarrow e\nu jj$ is given by
$[\epsilon_{WW}\cdot\sigma\cdot B(WW\rightarrow e\nu jj) +
\epsilon_{WZ}\cdot\sigma\cdot B(WZ\rightarrow e\nu jj)]/
[\sigma\cdot B(WW\rightarrow e\nu jj) +
\sigma\cdot B(WZ\rightarrow e\nu jj)] = (13.7 \pm 0.7)\%$, where
the theoretical cross sections of 9.5 pb for $WW$ and 2.5 pb for
$WZ$ production~\cite{Ohnemus}, and the $W$ and $Z$ boson branching fractions
from the Particle Data Group~\cite{PDG}, are used in the calculation
($\sigma\cdot B(WW\rightarrow e\nu jj)=1.38\pm0.05$ pb and
 $\sigma\cdot B(WZ\rightarrow e\nu jj)=0.188\pm0.006$ pb).

For the fast simulation, we generated over 30,000 events, with
approximately four times more for $WW$ production than
$WZ$ production, reflecting the sizes of their expected
production cross sections.
The overall detection efficiencies for
the SM couplings were calculated as
$[14.7 \pm 0.2 ({\rm stat}) \pm 1.2 ({\rm sys})]$\% for
$WW\rightarrow e\nu jj$ and $[14.6 \pm 0.4 ({\rm stat}) \pm 1.1 ({\rm sys})]$\%
for $WZ\rightarrow e\nu jj$.
The 7.8\% systematic uncertainty includes statistics of the
fast MC (1\%), efficiency of trigger and electron identification (1\%),
$\MET$ smearing and
modeling of the $p_T$ of the $WW/WZ$ system (5\%), difference in
$W\rightarrow jj$ detection efficiencies from the two event
generators (5\%),
and the effect of the jet energy scale (3\%).
The combined
efficiency is $[14.7 \pm 0.2 ({\rm stat}) \pm 1.2 ({\rm sys})]$\%.
The combined efficiency estimated using the fast
simulation is consistent with the value obtained using {\sc isajet}.

\subsection{Expected Number of Signal Events}
\label{expected-events}

Using the fast detector simulation and the cross section times branching ratio
from the event generator of Ref.~\cite{Hagiwara90} 
($\sigma\cdot B(WW \rightarrow e\nu jj)$ = 1.26 $\pm$ 0.18 pb, and
$\sigma\cdot B(WZ \rightarrow e\nu jj)$ = 0.18 $\pm$ 0.03 pb),
we estimate the number of expected $WW/WZ\rightarrow e\nu jj$
events to be 17.5 $\pm$ 3.0
(15.3 $\pm$ 3.0 $WW$ events and 2.2 $\pm$ 0.5 $WZ$ events), with
the uncertainty
(17.1\%) given by the sum in quadrature of the uncertainty in the efficiency,
the uncertainty in the luminosity (5.4\%), and that in the NLO
calculation (14\%).

\section{Background}
\label{background}

The sources of background to the 
$WW/WZ\rightarrow e\nu jj$ process
can be divided into two categories.
The first is instrumental background due to misidentified or
mismeasured particles,
and the other is inherent irreducible background consisting of physical
processes with the same signature as the events of interest.

\subsection{Instrumental Background}
\label{instrumental-background}

        The major source of instrumental background is QCD
multijet production in which one of the jets showers (mainly) in the
electromagnetic calorimeter and is misidentified as an electron, and
the energies of the remaining jets fluctuate to produce
$\MET$.
Although the probability for a jet to be misidentified as an electron
is small, the large cross section for QCD multijet
events makes this background significant. 

This background is estimated using samples of
``good" and ``bad" electrons.
A ``good" electron has the quality
requirements
described in Sec.~\ref{electron-selection}, while a ``bad"
electron has an EM cluster with
EM fraction $>$ 0.95, Isolation $\leq$ 0.15, and either $H$-matrix
$\chi^2$ $\geq$ 250  or track-match significance $\geq$ 10.
We assume that the shape of the $\MET$ spectrum of the events 
with a bad electron
is identical to the $\MET$ spectrum of the QCD multijet background.
Furthermore, with the assumption
that the contribution of signal events at low $\MET$
is negligible, the bad-electron sample can be
normalized to the good-electron
data in the low-$\MET$ region and the $\MET$ distribution
of the bad-electron events can then be
extrapolated to the signal region of the
good-electron sample.

\begin{figure}[htb]
\epsfxsize=3.4in
\centerline{\epsffile{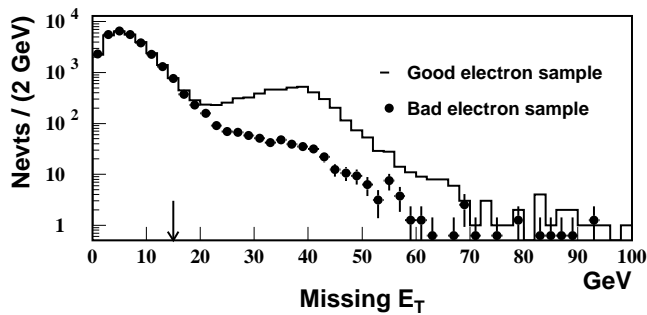}}  
\caption{$\MET$ distributions for the good-electron (histogram) and
bad-electron (solid circles) samples 
selected from data taken
with the {\tt  EM1\_ELE\_MON} and {\tt ELE\_1\_MON} triggers (see text).
The bad-electron sample is normalized to the good electron sample for
$\MET<15$ GeV.} 
\label{fig:met_mon}
\end{figure}

To estimate the multijet background, we use triggers that do not require
$\MET$.
Several L2 triggers in Run 1b meet this
requirement, in particular
the triggers {\tt EM1\_ELE\_MON} and {\tt ELE\_1\_MON}
described in Sec.~\ref{data-sample}.
To avoid biases, we add a condition that the EM object in these triggers
pass the same L2 requirements as the signal.
We then extract two
samples from these data, based on the electron quality. 
The $\MET$ distribution
for the bad-electron sample is then normalized to agree with the $\MET$
distribution for the good-electron sample at low $\MET$
($\MET<15$ GeV). Figure~\ref{fig:met_mon} shows these two distributions.
The normalization factor $N_{F}$ is calculated as the
ratio of the number of bad-electron events to
the number of good-electron events 
with $0\leq$~$\MET$~$\leq 15$~GeV. After imposing the
jet selection requirements on the events,
we find $N_{F}$ = 1.870 $\pm$ 0.060 (stat) $\pm$ 0.003 (sys).
The systematic uncertainty on the
normalization factor is obtained by varying the range of $\MET$ used for the
normalization procedure from 0--12 GeV to 0--18 GeV.

\begin{figure}[htb]
\epsfxsize=3.4in
\centerline{\epsffile{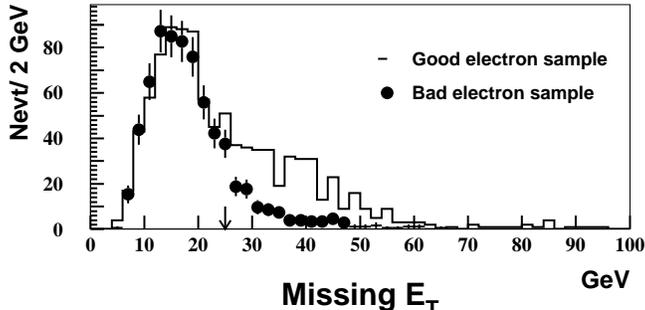}}
\caption{Distributions of $\MET$ of the good-electron (sum of signal
and background)
and bad-electron (background only) samples selected from data taken with
the trigger used for signal events.}
\label{fig:met_2j}
\end{figure}

In the next step, we select two samples from the data taken with the
trigger for signal events, one containing background and signal
(``good" electrons obtained through our selection procedure) 
and the other containing only background events (``bad" electrons). The
normalization factor $N_F$ is then applied to
the background sample.
Figure~\ref{fig:met_2j} shows the distributions of $\MET$ for the candidates and
the estimated QCD multijet background based on the bad-electron events
after the imposition of jet requirements.

From the above procedure, we estimate 104.3 $\pm$
8.2~(stat) $\pm$ 9.1~(sys) background events for $\MET>25$ GeV. 
The systematic uncertainty
(8.7\%)  includes the uncertainty on the normalization factor (1\%),
the difference when an alternative method is used to
estimate the multijet background (5.2\%), and
the difference for events with $\MET$ $> 25$ GeV
when the $\MET$ region 15--25 GeV
is used for normalization (6.9\%). 
In the alternative method, the probability of a jet to be misidentified as an
electron is multiplied by the number of multijet events
that satisfy selection criteria
when one of the jets in the event is treated as an electron.
When more than one jet in an event satisfies the kinematic requirements,
all are considered in estimating the background from multijet production.

\subsection{Inherent Background}
\label{physics-background}

The background contribution from processes with similar event topology
(i.e.,
with final-state objects identical to those of the signal) is estimated using
Monte Carlo events. 

\subsubsection{$W + \geq$ 2 jets}
\label{qcd-background}

\begin{figure}[htb]
\epsfxsize=3.4in
\centerline{\epsffile{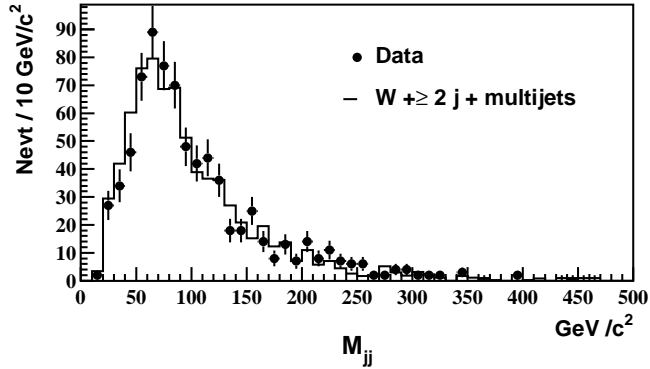}}  
\caption{Dijet invariant mass distribution. The solid circles and the
histogram are the candidate events and the background events
from $W+\ge 2$ jet events and QCD multijet events with a false electron,
respectively.}
\label{fig:mjj_data}
\end{figure}

\begin{figure}[htb]
\epsfxsize=3.4in
\centerline{\epsffile{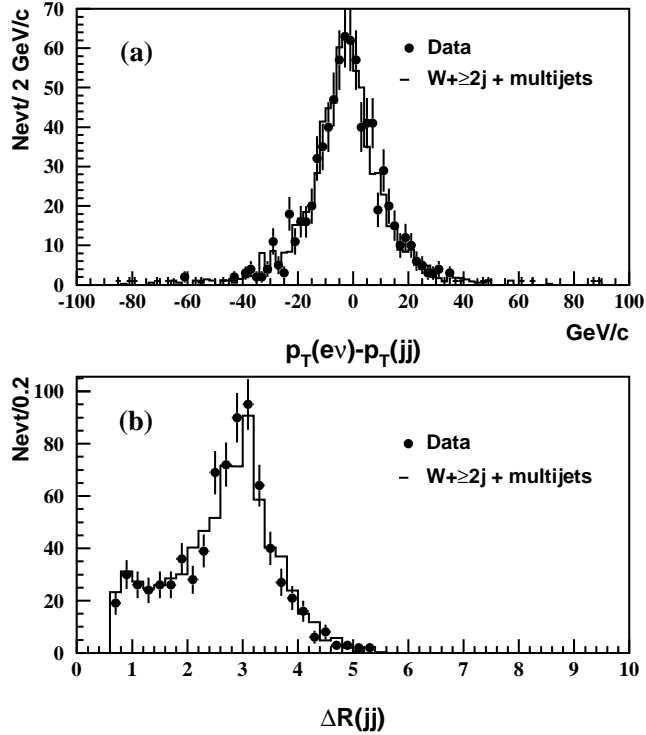}}
\caption{(a) Distributions in $p_T(e\nu) - p_T(jj)$ before imposition of the
mass window on $M_{jj}$. (b) Distributions for the separation of two jets
in $\eta-\phi$ space.}
\label{fig:key_dis}
\end{figure}

$W + \geq$ 2 jets production is the dominant background
to the $WW$ and $WZ$ signals.
This background is estimated using the Monte Carlo
program {\sc vecbos}~\cite{vecbos},
followed by {\sc herwig}~\cite{herwig} for the
hadronization of the partons generated in {\sc vecbos}
and then by the detailed simulation of the D{\O}
detector. The cross section from {\sc vecbos} 
has a large uncertainty, and the generated $W+\geq 2$ jets sample
is therefore normalized to the candidate
event sample after subtraction of the QCD multijet background.
To avoid the inclusion of $WW$ and $WZ$ events in this normalization
procedure, we use only the events whose two-jet invariant mass lies
outside of the mass peak of the $W$ boson (i.e., $M_{jj}>50$ or
$M_{jj}>110$ GeV/$c^2$).  Figure~\ref{fig:mjj_data} shows the two-jet
invariant mass distributions for data and the estimated background. The
normalization factor is found to be
$N_{V} = N_{VB}/(N_{\rm cand} - N_{\rm QCD}/N_{F})
= 3.41 \pm 0.31 ({\rm stat}) \pm 0.29 ({\rm syst})$,
where $N_{VB}=879$ corresonds to the number of {\sc vecbos} events,
$N_{\rm cand}=392$ is the number of candidate events in the data,
and $N_{\rm QCD}=251$ 
is the number of QCD multijet events outside of the $W$ boson mass window.
Using this normalization factor, we estimate 279.5 $\pm$ 27.2
(stat) $\pm$ 23.8 (sys) $W+ \geq 2$~jets
events in the candidate sample. The systematic
uncertainty is due to the normalization of the  multijet background (6.9\%),
uncertainty in the jet energy scale (4\%), and the difference observed
when the range of excluded $M_{jj}$ is changed to
40--120 or 60--100  GeV/$c^2$ (3\%).  The cross section multiplied by
the branching fraction 
for $W+\geq 2$~jets production, with the $W$ boson decaying to $e\nu$,
determined with this method is
$38795/(3.4 \times 82.3)= 138.6 \pm 14.3$ pb (where
38795 is the number of {\sc vecbos} events generated, 3.4 is the
normalization factor $N_{V}$, and 82.3 pb$^{-1}$
is the integrated luminosity 
of the data sample),
which is consistent with the value (135 pb) given by the {\sc vecbos}
program. Figure~\ref{fig:key_dis} shows distributions in the difference
$p_T(e\nu)-p_T(jj)$ and in the separation between jets
$\Delta {\cal R}(jj)$, which provide sensitive measures for how well
background estimates describe the jets in the data.
The backgrounds from the $W+\geq 2$ jets and QCD multijet
contributions are seen to agree well with the data.

\subsubsection{$t\overline{t}\rightarrow W^+W^-b\overline{b}\rightarrow e\nu
jj X$}
\label{top-background}

Since no limit on the number of
jets is applied to retain high efficiency,
$t\overline{t}\rightarrow W^+W^-b\overline{b}\rightarrow e\nu jj X$
events contribute to the candidate sample.
A sample, simulated using {\sc
isajet} with $M_t$ = 170 GeV/$c^2$, is used to estimate this
contribution.  We find it to be small, 3.7 $\pm$ 0.3 (stat) $\pm$ 1.3
(sys) events. The production cross section for $t{\bar t}$ events
is taken from the D\O\
measurement (5.2 $\pm$ 1.8 pb)~\cite{D0TopXsec}. The error in this
measurement (35\%) is included as a systematic uncertainty in our analysis.

\subsubsection{$WW/WZ\rightarrow \tau\nu jj\rightarrow e\nu\nu jj$}
\label{tau-backgrond}

    Since the contribution from
$WW/WZ\rightarrow \tau\nu jj\rightarrow e\nu\nu jj$ is small,
and no separate simulation of the signal 
is available, we treat it as background.
We use the {\sc isajet} event generator and the detailed detector simulation
program to estimate this source.
The $WW$ and $WZ$ production cross sections are assumed to be 9.5 pb and 2.5
pb, respectively.
After event selection,
we find 0.15~$^{+0.16}_{-0.08}$~(stat) $\pm$ 0.01~(sys)  events.
The systematic uncertainty on the background estimate is assigned to have the
larger value of
the asymmetric errors on the theoretical cross section (8.4\%)~\cite{Ohnemus}.

\subsubsection{$ZX\rightarrow e^+ e^- X$}
\label{z-background}

The $ZX \to ee X$ processes can produce events that can be
misidentified as signal. 
These events can be included in the candidate sample if one electron goes
through a boundary in a calorimeter module and is measured as $\MET$
in the event. From a sample of 10,000 {\sc isajet} $ZX\rightarrow
e^+ e^- X$ events generated, none survive the selection procedure. 
The background from events of this type is therefore negligible.  

\subsubsection{$ZX\rightarrow\tau^+ \tau^- X\rightarrow e\nu jj X$}
\label{ztau-background}

The $ZX \to \tau\tau X$ processes can also produce events that can be mistaken
for
signal if, due to shower fluctuation, one or two jets from
ISR or FSR are observed in the detector.  From a
sample of 10,000 {\sc pythia}-generated $ZX \to \tau\tau X$ events, none
survive our selection.  The background from this source is therefore also
negligible.

\section{Results}
\label{results}

\begin{figure}[htb]
\epsfxsize=3.4in
\centerline{\epsffile{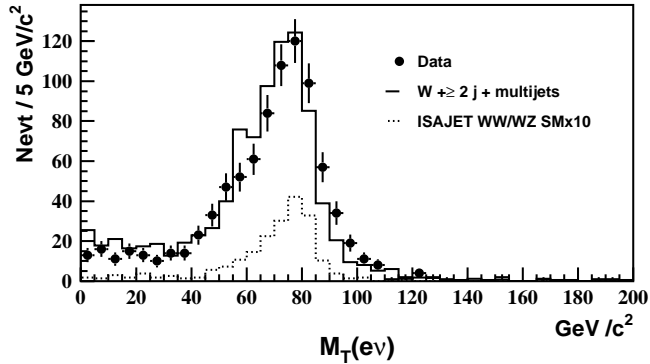}}  
\caption{Transverse-mass distributions of the electron and $\nu$ ($\MET$)
system. The solid circles, solid histogram, and dotted histogram are,
respectively, the candidate events, the background
from QCD multijet events with false electrons and $W+\ge 2$ jet events, and
the expected SM production of $WW/WZ$ events scaled up by a factor of ten.}
\label{fig:mtw}
\end{figure}

A total of 
399 candidate events remain after all selections.  The
number of events expected from SM $WW/WZ$ and
from SM background processes are $17.5\pm3.0$
and $387.5\pm38.1$, respectively.
The transverse mass distribution of the candidate events
is shown in Fig.~\ref{fig:mtw}, along with the contributions from background 
and the SM production of $WW/WZ$.
The distributions for data
agree well with expectations from background. 
Table~\ref{table:1a_1b} summarizes the
number of candidate events, the estimated backgrounds,
and SM predictions for the Run 1a and 1b data samples.

\begin{table}[htb]
\caption{Number of events for backgrounds, data and SM prediction
for Run 1a and Run 1b.}
\label{table:1a_1b}
\begin{center}
\begin{tabular}{ l  r@{$\pm$}l  r@{$\pm$}l }
                       &\multicolumn{2}{c}{Run 1a~\cite{D01aPRD}}
                       &\multicolumn{2}{c}{Run 1b}         \\ 
\hline
Luminosity             &\multicolumn{2}{c}{13.7 pb$^{-1}$}
                       &\multicolumn{2}{c}{82.3 pb$^{-1}$} \\ 
\hline
Background            &\multicolumn{2}{c}{}&\multicolumn{2}{c}{}\\ 
\hline   
QCD multijet          & 12.2 &\ 2.6 & 104.3 & 12.3 \\
$W + \geq $ 2 jets & 62.2 & 13.0 & 279.5 & 36.0 \\
$t{\bar t}  \rightarrow e\nu jj+X$ & 0.87 & 0.12 &   3.7 &\ 1.3 \\
\hline
Total background       & 75.5 & 13.3 & 387.5 & 38.1 \\
\hline
Data                   &\multicolumn{2}{c}{84}&\multicolumn{2}{c}{399}\\
\hline
$WW$+$WZ$ (SM prediction) & 3.2 &  0.6 &  17.5 &  3.0 \\
\end{tabular}
\end{center}
\end{table}

\begin{figure}[htb]
\epsfxsize=3.4in
\centerline{\epsffile{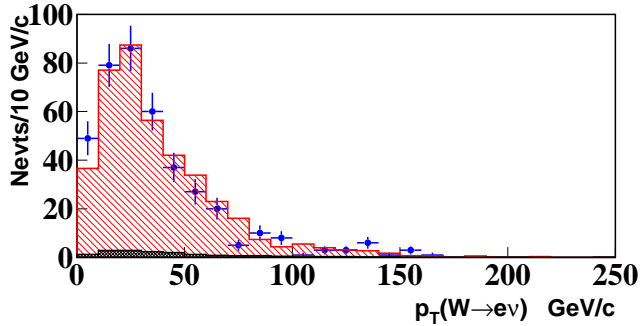}}
\caption{The $p_T$ distributions of the $e\nu$ system from the 1993--1995
(Run 1b) data. 
The solid circles are data.
The light-shaded
histogram is the SM prediction for the background, including the
dark-shaded histogram, which represents the SM prediction for
$WW/WZ$ processes.}
\label{fig:pt_all}
\end{figure}

	Figure~\ref{fig:pt_all} shows the $p_T$ distributions of the
$e\nu$ system for data, background estimates, and SM predictions.
We do not observe a statistically significant signal
above background.

Of the 399 events that satisfy the selection criteria, 18 events
have $p_T(e\nu) > 100$ GeV/$c$.
The numbers of background and SM events in this $p_T$ range
are estimated as $18.5\pm 1.8$ and $3.2\pm 0.5$, respectively.
The absence of an excess of
events with high $p_T(e\nu)$ excludes large deviations from SM couplings.

\subsection{Limits on Anomalous Couplings Using Minimum $p_T^W$}
\label{min-pt-limits}

The $WW/WZ$ production cross section  increases, especially at high $p_T^W$,
as the coupling parameters deviate from the SM values,
as shown in Fig.~\ref{fig:pt_anomalous}.
The $p_T^W$ distribution for background is softer than that of
$WW/WZ$ production with anomalous couplings.
When events are selected with $p_T^W$ above some large minimum value,
almost all background
events are rejected, but a good fraction of signal with anomalous
couplings remains, providing better sensitivity to such couplings.
This kind of selection eliminates most of SM production, and
therefore does not have sensitivity to the SM couplings. Moreover
the 95\% C.L. upper limit on the number of signal events
($N^{95\% {\rm C.L.}}$) can be obtained from
the observed number of candidate events and the expected background beyond
some minimum $p_T^W$ cutoff,
using the method described
in the report by the Particle Data Group~\cite{PDG}.
To do this,
Monte Carlo events are generated for pairs of anomalous
couplings in grid points of $\Delta\kappa$ and $\lambda$.
We assume that the couplings for $WW\gamma$
and $WWZ$ are equal.
The expected number of events for each pair of anomalous couplings
is calculated 
using the integrated luminosity of the data sample,
and entered into a two-dimensional density plot with
$\Delta\kappa$ and $\lambda$ as coordinate axes.
The results are fitted with a two-dimensional
parabolic function, and
limits on anomalous couplings are calculated at the 95\% C.L. from the
intersection of the two-dimensional parabolic surface for
the predicted number of
events with a plane of $N^{95\% {\rm C.L.}}$ values. 
The resulting contour of constant probability is an ellipse in the
$\Delta\kappa-\lambda$ plane.
The numerical values for the ``one-dimensional"
95\% C.L. limits (setting one of the coordinates to zero) are summarized in
Table~\ref{table:ptcutlim} for
different minimum values of $p_T^W$.

\begin{table}[htb]
\caption{Limits on $\lambda$ and $\Delta\kappa$ at the 95\% C.L. as a function
of minimum $p_T^W$, for $\Lambda$ = 1.5 TeV.
The number of candidates ($N_{\rm cand}$), background ($N_{\rm BG}$),
and the SM $WW/WZ$ predictions ($N_{\rm SM}$) are also listed.}
\label{table:ptcutlim}
\begin{center}
\begin{tabular}{ c  c  c  c  c  c }
$p^W_T$ (GeV/$c$)& $N_{\rm cand}$ & $N_{\rm BG}$ & $N_{\rm SM}$ &
$\lambda $ & $\Delta\kappa $ \\
& events& events& events& $(\Delta\kappa=0)$& $(\lambda=0)$\\
\hline
150 &  4 & 2.8 & 1.9 & $(-0.66,0.67)$ & $(-0.96,1.08)$\\
160 &  1 & 2.1 & 1.8 & $(-0.54,0.54)$ & $(-0.79,0.89)$\\
170 &  0 & 1.5 & 0.9 & $(-0.52,0.52)$ & $(-0.76,0.86)$\\
180 &  0 & 1.2 & 0.2 & $(-0.59,0.58)$ & $(-0.87,0.96)$\\
190 &  0 & 0.7 & 0.1 & $(-0.64,0.64)$ & $(-0.96,1.05)$\\
200 &  0 & 0.3 & 0.1 & $(-0.74,0.73)$ & $(-1.13,1.20)$\\
\end{tabular}
\end{center}
\end{table}

\subsection{Limits on Anomalous Couplings from the $p^W_T$ Spectrum}
\label{pt-fit-limits}

The limits obtained for some cutoff minimum $p_T^W$
do not take into account information that is available in
the full $p^W_T$ spectrum, and depend on the chosen minimum $p_T$ value as
well as on the overall normalization factors
for background and predictions for signal.
An alternative way to proceed
is to fit the shape of kinematical distributions that are
sensitive to anomalous couplings. This usually provides
tighter limits, since it uses all the information contained in the
differential distributions, and it is also less sensitive to overall
normalization factors.

As described in Sec.~\ref{introduction}, the
differential distribution that is most sensitive to anomalous couplings
is the $p_T^{W(Z)}$ distribution. Our analysis relies on the
$p_T(W\rightarrow e\nu)$ spectrum rather than $p_T(W\rightarrow jj)$ or
$p_T(Z\rightarrow jj)$ because the resolution on
$p_T(e\nu)$ (12.5 GeV/$c$) is better than on
$p_T(jj)$ (16.7 GeV/$c$).
This is primarily due to the ambiguity in assigning
jets to the $W(Z)$ boson.

	The differential cross sections 
have been exploited by previous publications
~\cite{CDFWg,D0Wg,D01aPRD,D0WW,CDFWZ,D0WZ1a,D0WZ1bmu,D0WZ1b,Sanchez-thesis}
for extracting limits on trilinear gauge boson couplings.
We use a modified fit to the binned $p_T^W$ distribution to obtain limits, with 
the modification consisting of adding an extra bin in $p^W_T$
with no observed events, thereby
improving the sensitivity to anomalous
couplings~\cite{Landsberg}.

Based on the number of expected $WW/WZ\rightarrow e\nu jj$ events, we choose
two 25 GeV/$c$ bins between 0 and 50 GeV/$c$, five 10 GeV/$c$ bins from 50 to 100
GeV/$c$, two 20 GeV/$c$ bins from 100 to 140 GeV/$c$, one 30 GeV/$c$
bin from 140
to 170 GeV/$c$, and a single bin from 170 GeV/$c$ to 500 GeV/$c$.
The cross section for $p_T^W>500$ GeV/$c$
is negligible for any anomalous couplings
allowed by unitarity.
For each bin $i$ of $p^W_T$, the
probability $P_i$ for observing $N_i$ events is given by the Poisson
distribution: 
$$ P_i = \frac{(b_i +
{\cal L}\epsilon_i\sigma_i(\lambda,\Delta\kappa))^{N_i}}{N_i !}e^{-(b_i +
{\cal L}\epsilon_i\sigma_i(\lambda,\Delta\kappa))},$$
where ${\cal L}$ is the luminosity, and $b_i$, $\epsilon_i$, and
and $\sigma_i$ are the expected background, 
the total detection efficiency, and
the cross section, respectively, for bin $i$.
Our fast Monte Carlo is used to calculate
$\epsilon_i\sigma_i(\lambda,\Delta\kappa)$.
The joint probability for all $p^W_T$ bins is the product
of the individual probabilities $P_i$,
$P = \prod^{N_{\rm bin}}_{i=1} P_i.$
Since the values ${\cal L}$, $b_i$, and $\epsilon_i$ are measured
values with their respective uncertainties, we assign them
Gaussian prior distributions of mean $\mu = 1$ and standard
deviation $\sigma_x$: 
$$P' = \int {\cal{G}}_{f_n} df_n \int {\cal{G}}_{f_b} df_b 
 \prod^{N_{\rm bin}}_{i=1} \frac{e^{f_n n_i + f_b b_i}(f_n n_i + f_b
b_i)^{N_i}}{N_i !},$$
where $n_i =
{\cal L}\epsilon_i\sigma_i$ is the predicted number of signal events, and 
${\cal{G}}_{f_n}$ and ${\cal{G}}_{f_b}$ are Gaussian distributions for the
fractions of signal and background events.
The integrals are calculated using 50 evenly spaced points
between $\pm3$ standard deviations.
For convenience, the log of the likelihood $L = \log P'$ is
used in the fit and
the set of couplings that best describes the data is given by the
point in the $\lambda-\Delta\kappa$ plane that maximizes the likelihood given
in the above equation.

	It is conventional to quote the
limits on one coupling when all the others are set to their SM
values. These ``one-dimensional" limits at the 
95\% C.L., assuming that the $WW\gamma$
and $WWZ$ couplings are equal,
are shown in Table~\ref{table:lim_1b}. 
The limits are more stringent
than those obtained using the minimum $p_T^W$ method. 

\begin{table}[htb]
\caption{Limits on anomalous trilinear gauge boson couplings at the
95\% C.L. for three values of $\Lambda$ obtained using the fit to $p_T^W$ 
for data from the Run 1b.}
\label{table:lim_1b}
\begin{center} 
\begin{tabular}{cccc}
Couplings & $\Lambda=1.0$ TeV & $\Lambda=1.5$ TeV & 
$\Lambda=2.0$ TeV \\
\hline
$\lambda_{\gamma} = \lambda_Z$ ($\Delta\kappa_{\gamma}=\Delta\kappa_Z=0$)&
$-0.50,0.53$ & $-0.42,0.45$ & $-0.39,0.42$\\
$\Delta\kappa_{\gamma}=\Delta\kappa_Z$ ($\lambda_{\gamma} = \lambda_Z=0$)&
$-0.66,0.90$ & $-0.56,0.75$ & $-0.52,0.70$\\
\hline
$\lambda_{\gamma}$ (HISZ) ($\Delta\kappa_{\gamma}=0$)&
$-0.50,0.53$ & $-0.42,0.45$ & $-0.39,0.42$\\
$\Delta\kappa_{\gamma}$ (HISZ) ($\lambda_{\gamma}=0$)&
$-0.78,1.15$ & $-0.68,0.98$ & $-0.63,0.91$\\
\hline
$\lambda_{\gamma}$ (SM $WWZ$) ($\Delta\kappa_{\gamma}=0$)&
$-1.54,1.58$ & $-1.53,1.56$&\\
$\Delta\kappa_\gamma$ (SM $WWZ$) ($\lambda_{\gamma}=0$)&
$-2.03,2.45$ & $-1.79,2.12$&\\
\hline
$\lambda_Z$ (SM $WW\gamma$) ($\Delta\kappa_Z=\Delta g^Z_1=0$)&
$-0.58,0.62$ & $-0.49,0.51$ & $-0.45,0.48$\\
$\Delta\kappa_Z$ (SM $WW\gamma$) ($\lambda_Z=\Delta g^Z_1=0$)&
$-0.86,1.12$ & $-0.72,0.93$ & $-0.67,0.87$\\
\end{tabular}
\end{center}
\end{table}

	We have assumed thus far that the couplings $\Delta\kappa$ and
$\lambda$ for $WWZ$ and $WW\gamma$ are equal. However, this is not the
only possibility.
Another common assumption leads to the HISZ relations~\cite{HISZ}. These
relations specify $\lambda_Z$, $\kappa_Z$, and $g_1^Z$ in terms of the
independent variables $\lambda_{\gamma}$ and $\kappa_{\gamma}$, thereby
reducing the number of independent couplings from five to two:
$\Delta\kappa_Z = \frac{1}{2}\Delta\kappa_{\gamma}(1-\tan^2 \theta_W)$,
$\Delta g_1^Z = \frac{1}{2}\Delta\kappa_{\gamma}/\cos^2 \theta_W$, and
$\lambda_Z = \lambda_{\gamma}$.
These one-dimensional limits at the 95\% C.L. are also shown in
Table~\ref{table:lim_1b}.

Since the $WWZ$ and $WW\gamma$ couplings are independent, it is
interesting to find the limits on one when the
other is set to its SM values. 
Table~\ref{table:lim_1b} includes the one-dimensional limits at the
95\% C.L. for both assumptions: limits on
$\Delta\kappa_{\gamma}$ and $\lambda_{\gamma}$ when the $WWZ$ couplings are
assumed to be standard, and
limits on $\Delta\kappa_{Z}$ and $\lambda_{Z}$ when the $WW\gamma$ couplings
are assumed to be standard. These results indicate that our analysis is more
sensitive to $WWZ$ couplings, as should be expected from the larger overall
SM couplings for $WWZ$ than for $WW\gamma$, and
that our analysis is complementary to studies of the $W\gamma$
production process which is sensitive only to the $WW\gamma$ couplings.

\subsection{Combined Results for Run 1 on $WW/WZ\rightarrow e\nu jj$}
\label{combined-limits}

	The limits on anomalous couplings presented in this
paper are significantly tighter than those in our previous
publications~\cite{D01aPRD,D0WZ1a}.
The primary reason for this is the
increase in the amount of data (about a factor of six).
We can obtain even stronger
limits by combining the results from Run 1a and 1b. The 
analysis based on the Run 1a data is described in
Refs~\cite{D01aPRD,D0WZ1a}. A summary of the signal and
backgrounds for the two analyses~\cite{D0WZ1b} is given in
Table~\ref{table:1a_1b}.

\begin{figure}[htb]
\epsfxsize=3.46in
\centerline{\epsffile{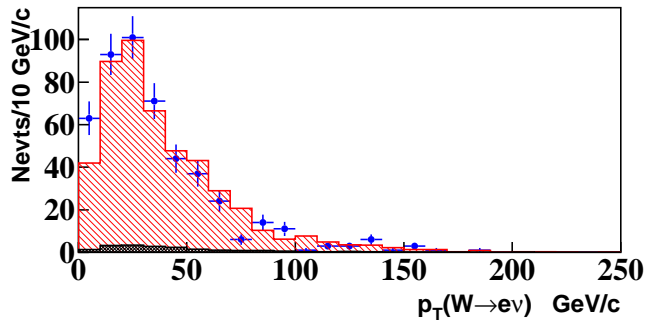}}
\caption{The $p_T^W$ spectrum for $e\nu jj$
candidates for the full Run 1 data sample.
The solid circles are data.
The light-shaded
histogram is the sum of predictions from the SM and background, and the
dark-shaded histogram is the SM prediction for $WW/WZ$ processes alone.}
\label{fig:pt_combine}
\end{figure}

The two analyses can be treated as different experiments. However,
because both experiments used the same detector, there are certain
correlated uncertainties, such as the uncertainties on the  
luminosity, lepton reconstruction and identification, and
the theoretical prediction. Also,
the background estimate is common to each experiment.
The uncertainty on the $W/Z\rightarrow jj$ selection efficiency
is assumed to be
uncorrelated, since we use different cone sizes for jet 
reconstruction in the two analyses.
(This hypothesis does not affect the results in any significant way.)
The uncertainties for both analyses are
summarized in Tables~\ref{table:common_err} and \ref{table:uncorrel_err}.
Each uncertainty is weighted by
the integrated luminosity for the respective data sample. 
Figure~\ref{fig:pt_combine} shows the combined $p_T^W$ spectrum.

\begin{table}[htb]
\caption{Common systematic uncertainties for Run 1a
and Run 1b analyses.}
\begin{center}
\begin{tabular}{lr}
Source of Uncertainty   &    \\ \hline
Luminosity           &5.4\% \\
QCD corrections                 &14\%  \\
Electron and trigger efficiency &1.2\% \\
Statistics of fast MC           &1\%   \\
$\MET$ smearing                 &5.1\% \\ 
Jet energy scale                &3.4\% \\ \hline
Total        &16\%\\
\end{tabular}
\label{table:common_err}
\end{center}
\end{table}

\begin{table}[htb]
\caption{Uncorrelated systematic uncertainties for Run 1a
and Run 1b analyses.}
\begin{center}
\begin{tabular}{lrr}
Source of Uncertainty   &   Run 1a    &   Run 1b \\ \hline
{\sc isajet} vs {\sc pythia}   &      9\%     &      4\%    \\
Stat. uncertainties of $\epsilon(W\rightarrow jj)$ & 4\%     &      2\%    \\
Parametrization of $\epsilon_i \sigma_i(\lambda,\Delta\kappa)$    &      4\%     &      5\%    \\
Total (added in quadrature)  &    11\%    &     7\%   \\ \hline
Background&      13\%    &       7\%   \\
\end{tabular}
\label{table:uncorrel_err}
\end{center}
\end{table}

	To set limits on anomalous couplings, we combine the
results of the two analyses by calculating a combined likelihood function.
The individual uncertainties on
signal and background for each analysis are taken into account as in the
previous section. Common systematic
uncertainties are taken
into account by introducing a common Gaussian prior distribution for the two
data samples.

\begin{figure}[htb]
 \begin{center}
 \begin{tabular}{cc}
    \epsfxsize=1.7in
    \epsffile{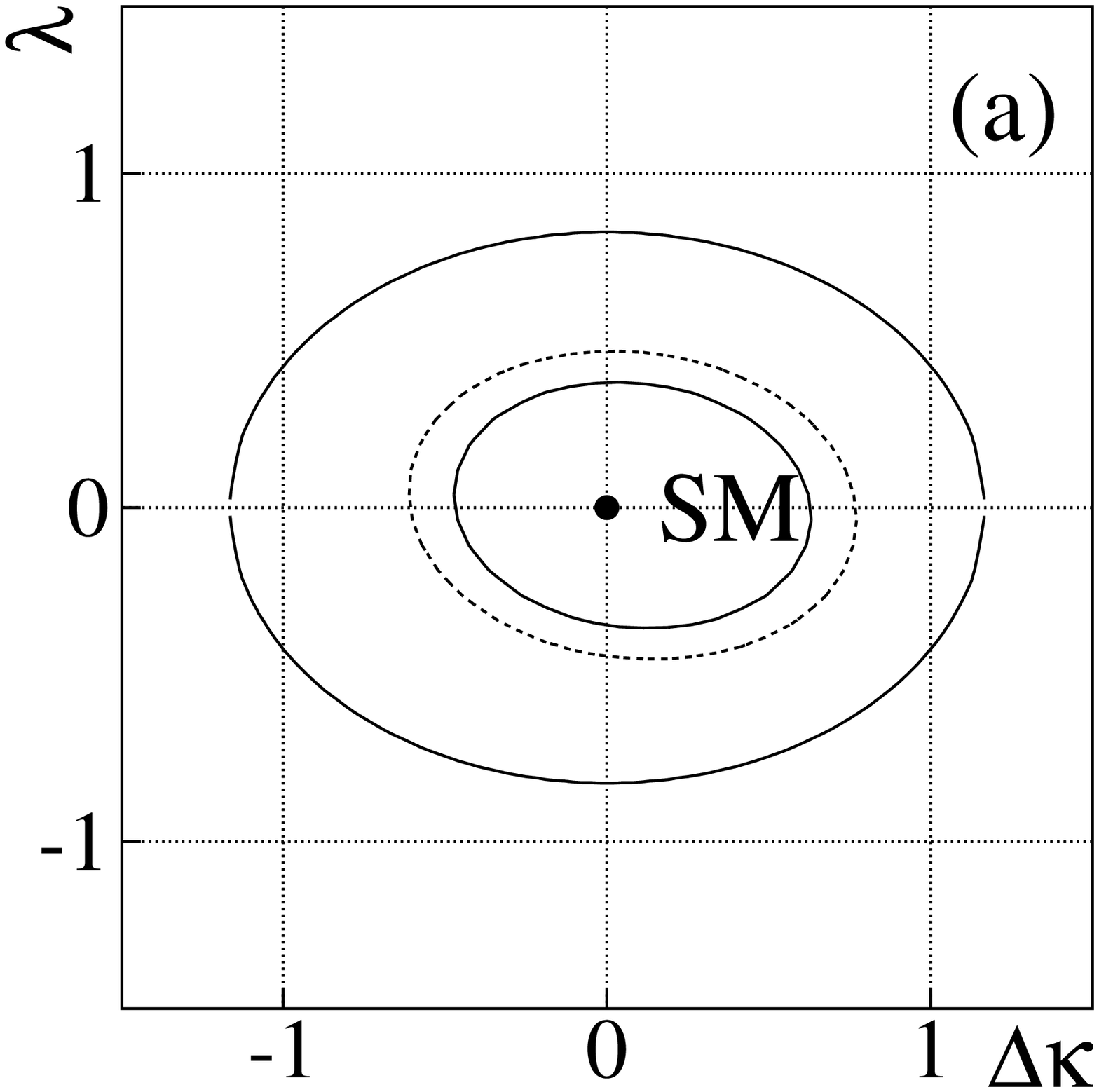}&
    \epsfxsize=1.7in
    \epsffile{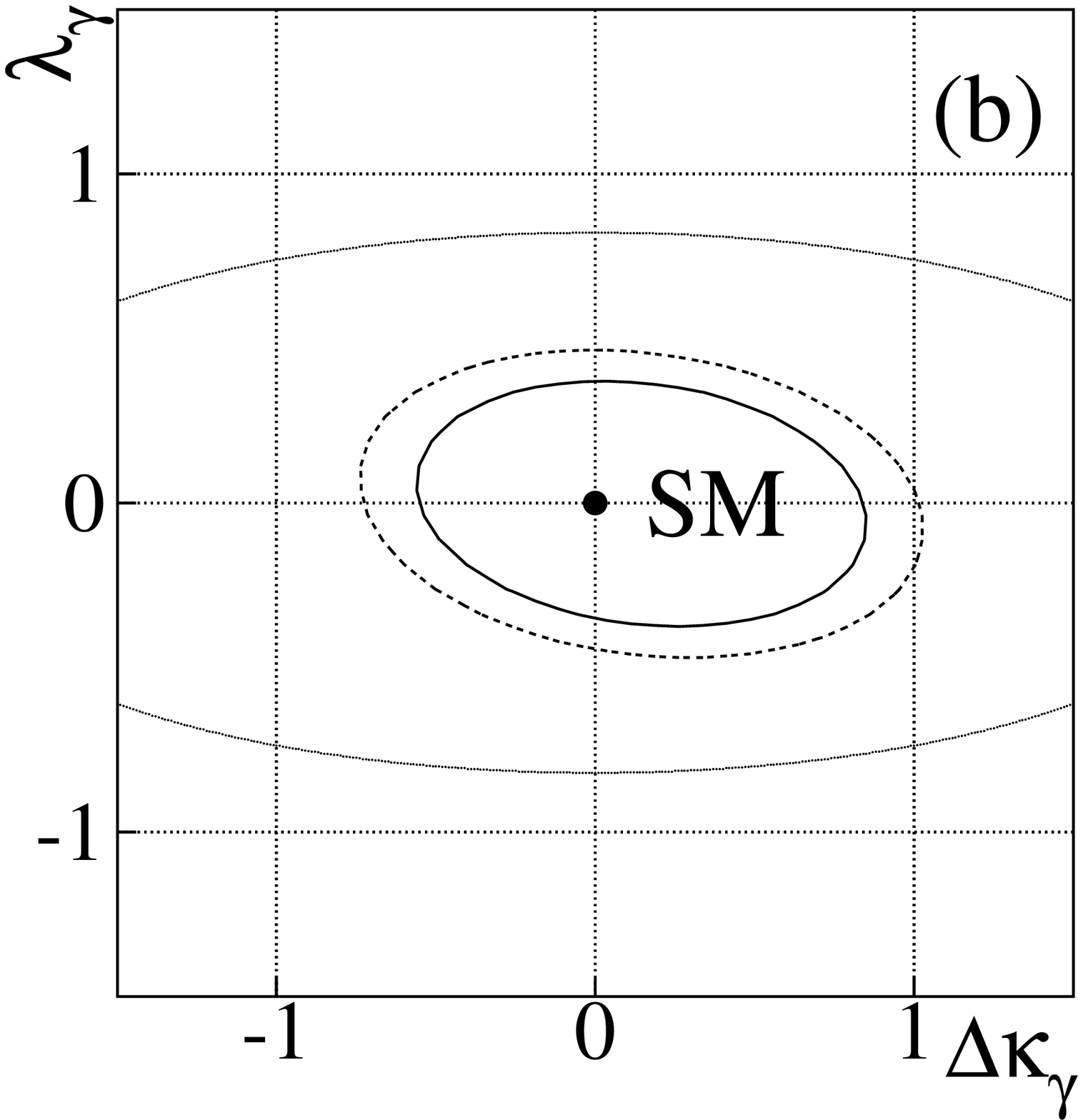}\\
    \epsfxsize=1.7in 
    \epsffile{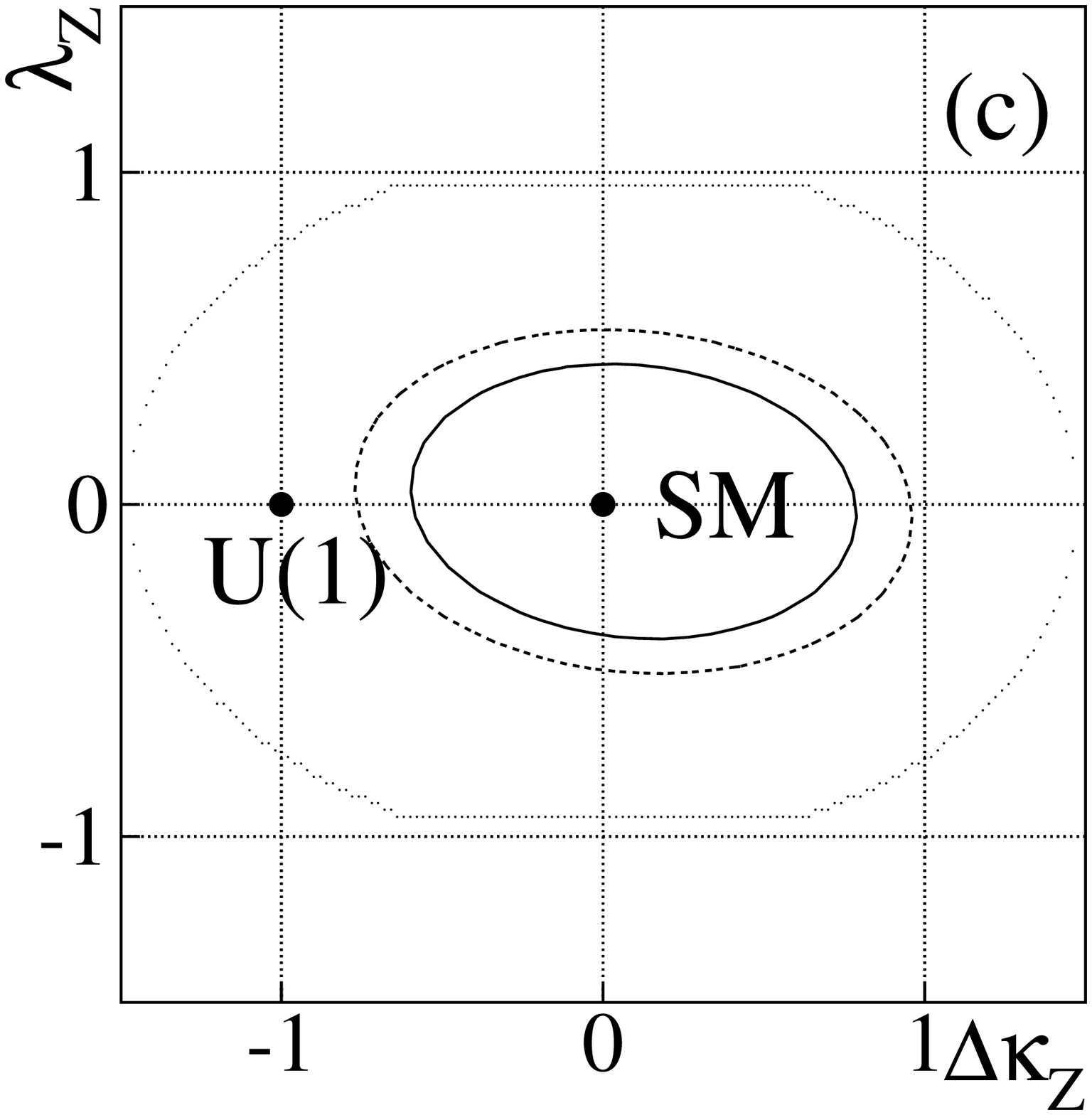}  &
    \epsfxsize=1.7in
    \epsffile{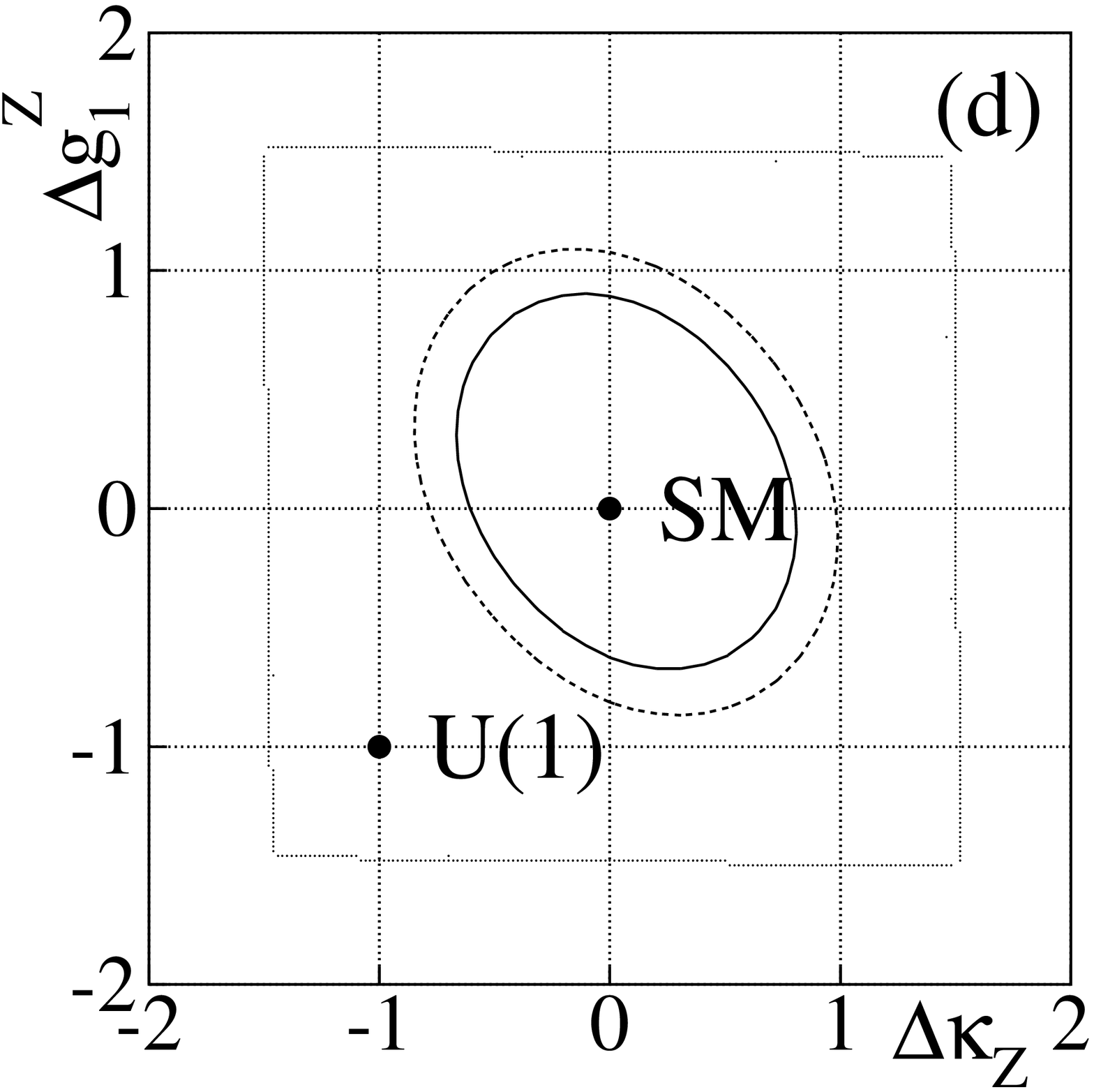}
 \end{tabular}
\caption{Contour limits on anomalous couplings at the 95\% C.L.
(two inner curves) and unitary constraints (outermost curves), assuming (a)
$\Delta\kappa \equiv \Delta\kappa_{\gamma} = \Delta\kappa_{Z},
    \lambda \equiv \lambda_{\gamma} = \lambda_{Z}$; (b) HISZ relations;
(c) and (d) SM $WW\gamma$ couplings. 
$\Lambda = 1.5$ TeV is used for all four cases.
The U(1) point is the expectation with no $WWZ$ couplings.}
\label{fig:all_com}
 \end{center}
\end{figure}

	Combining results from Run 1a and Run 1b yields the 95\% C.L. contours
of constant probability shown in Fig.~\ref{fig:all_com}.
The one and two dimensional 95\% C.L. contour limits (corresponding
to log-likelihood values of 1.92 and 3.00 units below the maximum,
respectively) are shown as the inner contours, along with the
unitarity limits from the $S$-matrix, shown as the outermost contours.
Figure~\ref{fig:all_com}(a) shows the contour limits when couplings
for $WW\gamma$ are assumed to be equal to those for $WWZ$.
Figure~\ref{fig:all_com}(b) shows contour limits assuming the HISZ relations.
In Figs.~\ref{fig:all_com}(c) and \ref{fig:all_com}(d), SM $WW\gamma$ couplings
are assumed and the limits are shown for $WWZ$ couplings.
Assuming SM $WW\gamma$ couplings, the U(1) point
that corresponds to the condition in which there is no $WWZ$ couplings
($\kappa_Z=0$, $\lambda_Z=0$, $g_1^Z=0$) is excluded at the 99\% C.L.
This is direct evidence for the existence of $WWZ$ couplings.
These limits are slightly stronger than those
from the 1993--1995 data alone.
The one-dimensional 95\% C.L. limits
for four assumptions on the relation between
$WW\gamma$ and $WWZ$ couplings: (i) $\Delta\kappa\equiv\Delta\kappa_\gamma =
\Delta\kappa_Z$, $\lambda\equiv\lambda_\gamma=\lambda_Z$, (ii) HISZ relations,
(iii) SM $WW\gamma$ couplings, and (iv) SM $WWZ$ couplings
are listed in Table~\ref{table:lim_com}.

\begin{table}[htb]
\caption{Limits on anomalous trilinear gauge boson couplings at 95\% C.L.
from the combined Run 1a and Run 1b data samples for these values of $\Lambda$.}
\label{table:lim_com}
\begin{center}
\begin{tabular}{cccc}
Couplings & $\Lambda=1.0$ TeV & $\Lambda=1.5$ TeV & 
$\Lambda=2.0$ TeV \\
\hline
$\lambda_{\gamma}=\lambda_Z$ ($\Delta\kappa_{\gamma}=\Delta\kappa_Z=0$)&
$-0.42,0.45$ & $-0.36,0.39$ & $-0.34,0.36$\\
$\Delta\kappa_{\gamma}=\Delta\kappa_Z$ ($\lambda_{\gamma}=\lambda_Z=0$)&
$-0.55,0.79$ & $-0.47,0.63$ & $-0.43,0.59$\\
\hline
$\lambda_{\gamma}$ (HISZ) ($\Delta\kappa_{\gamma}=0$)&
$-0.42,0.45$ & $-0.36,0.39$ & $-0.34,0.36$\\
$\Delta\kappa_{\gamma}$ (HISZ) ($\lambda_{\gamma}=0$)&
$-0.69,1.04$ & $-0.56,0.85$ & $-0.53,0.78$\\
\hline
$\lambda_{\gamma}$ (SM $WWZ$) ($\Delta\kappa_\gamma=0$)&
$-1.28,1.33$& $-1.21,1.25$&\\
$\Delta\kappa_\gamma$ (SM $WWZ$) ($\lambda_{\gamma}=0$)&
$-1.60,2.03$& $-1.38,1.70$&\\
\hline
$\lambda_Z$ (SM $WW\gamma$) ($\Delta\kappa_Z=\Delta g_1^Z=0$)&
$-0.47,0.51$ & $-0.40,0.43$ & $-0.37,0.40$\\
$\Delta\kappa_Z$ (SM $WW\gamma$) ($\lambda_Z=\Delta g_1^Z=0$)&
$-0.74,0.99$ & $-0.60,0.79$ & $-0.54,0.72$\\
$\Delta g_1^Z$ (SM $WW\gamma$) ($\lambda_Z=\Delta\kappa_Z=0$)&
$-0.75,1.06$ & $-0.64,0.89$ & $-0.60,0.81$\\
\end{tabular}
\end{center}
\end{table}

\section{Conclusions}
\label{conclusions}

	We have searched for anomalous $WW$ and $WZ$ production in the $e\nu jj$
decay mode at $\sqrt{s} = 1.8$ TeV. In a total of
82.3 pb$^{-1}$ of data
from the 1993--1995 collider run at Fermilab, we observe 399
candidate events with an expected background of 387.5$\pm$38.1 events.
The expected number of events from SM $WW/WZ$ production is $17.5\pm3.0$ 
events for
this integrated luminosity. The sum of the SM prediction and the
background estimates is consistent with the observed number of events,
indicating that no new physics phenomena is seen.
Comparing the $p_T^W$ distributions of the observed events
with theoretical predictions, we set limits on the
$WW\gamma$ and $WWZ$ anomalous couplings.
The limits on anomalous couplings are significantly tighter than those
using the 1992--1993 data sample.
The two results are combined to set
even tighter limits on the anomalous couplings.
With an assumption that the $WW\gamma$ and $WWZ$ couplings are equal, we obtain
$-0.34 < \lambda < 0.36$ (with $\Delta\kappa=0$) and
$-0.43 < \Delta\kappa < 0.59$ (with $\lambda=0$) at the 95\% C.L. for a form
factor scale $\Lambda=2.0$ TeV~\cite{Combined_Fit}.

\section{Acknowledgements}
%
We thank the Fermilab and collaborating institution staffs for 
contributions to this work, and acknowledge support from the 
Department of Energy and National Science Foundation (USA),  
Commissariat  \` a L'Energie Atomique (France), 
Ministry for Science and Technology and Ministry for Atomic 
   Energy (Russia),
CAPES and CNPq (Brazil),
Departments of Atomic Energy and Science and Education (India),
Colciencias (Colombia),
CONACyT (Mexico),
Ministry of Education and KOSEF (Korea),
CONICET and UBACyT (Argentina),
A.P. Sloan Foundation,
and the Humboldt Foundation.
%


\end{document}